\documentclass[referee]{aa} 
\usepackage{graphicx}
\usepackage{txfonts}
\usepackage{bm}

\def\be{\begin{equation}}
\def\ee{\end{equation}}
\def\bea{\begin{eqnarray}}
\def\eea{\end{eqnarray}}
\def\btable{\begin{table}}
\def\etable{\end{table}}
\def\btabular{\begin{tabular}}
\def\etabular{\end{tabular}}
\def\bdoc{\begin{document}}
\def\edoc{\end{document}}
\def\bfig{\begin{figure}}
\def\efig{\end{figure}}
\def\ben{\begin{enumerate}}
\def\een{\end{enumerate}}
\def\bit{\begin{itemize}}
\def\eit{\end{itemize}}
\def\bitem{\begin{itemize}}
\def\eitem{\end{itemize}}
\def\bquote{\begin{quote}}
\def\equote{\end{quote}}

\def\bcenter{\begin{center}}
\def\ecenter{\end{center}}

\def\bsmall{\begin{small}}
\def\esmall{\end{small}}
\def\blarge{\begin{large}}
\def\elarge{\end{large}}
\def\bLarge{\begin{Large}}
\def\eLarge{\end{Large}}
\def\bhuge{\begin{huge}}
\def\ehuge{\end{huge}}
\def\bHuge{\begin{Huge}}
\def\eHuge{\end{Huge}}
\def\bfoot{\begin{footnotesize}}
\def\efoot{\end{footnotesize}}

\def\n{\noindent}
\def\bb{\left(}
\def\eb{\right)}
\def\bbs{\left[}
\def\ebs{\right]}
\def\bba{\left<}
\def\eba{\right>}
\def\bbsq{\left[}
\def\ebsq{\right]}
\def\bbcu{\left\{}
\def\ebcu{\right\}}
\def\bml{\left|}
\def\emr{\right|}
\newcommand \sib [1] {\bb #1 \eb}

\newcommand \sqb [1] {\bbsq #1 \ebsq}
\newcommand \cub [1] {\bbcu #1 \ebcu}
\def\etal{et al.\ }
\def\ie{i.\,e.\,, }
\def\eg{e.\,g.\,} 
\def\etc{etc.\ }
\def\viz{viz.\,, }

\def\gne #1#2{\ \vphantom{S}^{\raise-0.5pt\hbox{$\scriptstyle#1$}}_
{\raise0.5pt \hbox{$\scriptstyle#2$}}}
\def\gneq{\gne > \sim}
\def\lneq{\gne < \sim}

\def\mod #1{|#1|}
\def\vec#1{{\bf #1}}
\def\dotvec#1{\dot\vec{#1}}
\def\mvec#1{\mod{\vec{#1}}}
\def\mean #1{\bba #1 \eba}
\def\meanvsq#1{\mean{\vec{#1}^2}}
\def\dotp#1#2{\vec{#1}\cdot\vec{#2}}
\def\meandotp#1#2{\mean{\dotp{#1}{#2}}}
\def\crossp#1#2{\vec{#1}\times\vec{#2}}
\def\vecp{\vec{p}}
\def\vecpone{\vec{p}_{1}}
\def\vecn{\vec{n}}
\def\vecnone{\vec{n}_{1}}

\def\reci #1{\frac {1} {#1}}
\def\ten#1#2{#1{\times10^{#2}}}
\def\onlyten#1{10^{#1}}
\def\deg{\unit{deg}}
\def\degsq{\unit{deg^2}}
\def\ster{\unit{steradian}}
\def\perc{\unit{percent}}
\def\percent{\unit{percent}}

\def\deri#1#2{\frac {d#1} {d#2}}
\def\deriemp#1{\frac {d} {d#1}}
\def\pderi#1#2{\frac {\partial#1} {\partial#2}}
\def\pderiemp#1{\frac {\partial} {\partial#1}}
\def\secderi#1#2{\frac {d^2#1} {d#2^2}}
\def\secpderi#1#2{\frac {\partial^2#1} {\partial#2^2}}

\def\grad{\nabla}

\def\unit #1{\,{\rm #1}} 

\def\angst{\unit{\AA}}

\def\gm{\unit{gm}}
\def\gmi{\gm^{-1}}
\def\mg{\unit{mg}}
\def\kg{\unit{kg}}
\def\nm{\unit{nm}}
\def\mum{\unit{\mu m}}
\def\mumi{\unit{\mu\,m^{-1}}}
\def\mm{\unit{mm}}
\def\cm{\unit{cm}}
\def\mtr{\unit{m}}
\def\mtri{\unit{m^{-1}}}
\def\km{\unit{km}}

\def\pc{\unit{pc}}
\def\kpc{\unit{kpc}}
\def\mpc{\unit{Mpc}}
\def\gpc{\unit{Gpc}}

\def\cmi{\unit{cm^{-1}}}
\def\cmsq{\unit{cm^2}}
\def\cmsqi{\unit{cm^{-2}}}
\def\cmcu{\unit{cm^3}}
\def\cmcui{\unit{cm^{-3}}}
\def\kmi{\unit{km^{-1}}}
\def\pci{\unit{pc^{-1}}}
\def\kpci{\unit{kpc^{-1}}}
\def\mtrsq{\unit{m^2}}
\def\mtrsqi{\unit{m^{-2}}}
\def\mtrcu{\unit{m^3}}
\def\mtrcui{\unit{m^{-3}}}
\def\mpccu{\unit{cm^3}}
\def\mpccui{\unit{cm^{-3}}}
\def\gpccu{\unit{cm^3}}
\def\gpccui{\unit{Gpc^{-3}}}
\def\kpsi{\unit{km \, s^{-1}}}

\def\ksec{\unit{ksec}}
\def\msec{\unit{msec}}
\def\musec{\unit{\mu\,sec}}
\def\min{\unit{min}}
\def\hr{\unit{hr}}
\def\day{\unit{day}}
\def\days{\unit{days}}
\def\week{\unit{week}}
\def\yr{\unit{yr}}
\def\myr{\unit{Myr}}
\def\gyr{\unit{Gyr}}
\def\micron{\unit{\mu m}}
\def\delto{\Delta t_{\rm o}}

\def\seci{\unit{s^{-1}}}
\def\dayi{\unit{day^{-1}}}
\def\yri{\unit{yr^{-1}}}

\def\cms{\unit{cm\,s^{-1}}}
\def\mts{\unit{m\,s^{-1}}}
\def\mtss{\unit{m\,s^{-2}}}
\def\kms{\unit{km\,s^{-1}}}
\def\kmh{\unit{km\,h^{-1}}}

\def\mas{\unit{mas}}
\def\muas{\unit{{\mu}}as}
\def\masyr{\unit{mas\,yr^{-1}}}
\def\arcs{\unit{arcsec}}
\def\arcsec{\unit{arcsec}}
\def\arcsecsq{\unit{arcsec^2}}
\def\arcsecsqi{\unit{arcsec^{-2}}}
\def\arcmin{\unit{arcmin}}
\def\arcm{\unit{arcmin}}
\def\arcmsq{\unit{arcmin^2}}
\def\degsq{\unit{deg^2}}
\def\degsqi{\unit{deg^{-2}}}

\def\vlos{\vec{v}_{\rm los}}

\def\hz{\unit{Hz}}
\def\khz{\unit{kHz}}
\def\mhz{\unit{MHz}}
\def\ghz{\unit{GHz}}
\def\thz{\unit{Thz}}
\def\ev{\unit{eV}}
\def\kev{\unit{keV}}
\def\mev{\unit{MeV }}
\def\gev{\unit{GeV }}
\def\tev{\unit{TeV }}

\def\kel{\unit{K}}

\def\newt{\unit{N}}

\def\funit{\rm\,erg\,cm^{-2}\,sec^{-1}}
\def\pfunit{\unit{cm^{-2}\,sec^{-1}}}
\def\funith{\rm\,erg\,cm^{-2}\,sec^{-1}\,Hz^{-1}}
\def\funitb{\rm\,erg\,cm^{-2}\,sec^{-1}\ster^{-1}}
\def\funitbh{\rm\,erg\,cm^{-2}\,sec^{-1}\Hz^{-1}\ster^{-1}}
\def\funitba{\rm\,erg\,cm^{-2}\,sec^{-1}\AA^{-1}\ster^{-1}}
\def\punitb{\rm\,photons\,cm^{-2}\,sec^{-1}\ster^{-1}}
\def\lunit{\rm\,erg\,sec^{-1}}
\def\lunith{\rm\,erg\,sec^{-1}\,Hz^{-1}}
\def\whz{\unit{W\,Hz^{-1}}}
\def\whzs{\unit{W\,Hz^{-1}\,str^{-1}}}
\def\wmtrsqi{\unit{W\,m^{-2}}}
\def\jy{\unit{Jy}}
\def\mjy{\unit{mJy}}
\def\mujy{\unit{\mu Jy}}
\def\jyu{\unit{W\,m^{-2}\,Hz}}
\def\erg{\unit{erg}}
\def\ergs{\unit{erg\,s^{-1}}}
\def\ergsh{\unit{erg\,s^{-1}\,Hz^{-1}}}
\def\mcrab{\unit{milliCrab}}
\def\ergsc{\unit{erg\,s^{-1}\,cm^{-3}}}
\def\ergcmcu{\unit{erg\,cm^{-3}}}
\def\ergscmcu{\unit{erg\,s^{-1}\,cm^{-3}}} 
\def\ergcmt{\unit{erg\,cm^{-3}}}
\def\ergcmcui{\unit{erg\,cm^{-3}}}

\def\gmcmcui{\gm\cmcui}
\def\kgmtrcui{\kg\mtrcui}
 
\def\melec{m_{\rm e}}
\def\mectwo{\melec c^2}
\def\mprot{m_{\rm p}}
\def\nelec{n_{\rm e}}
\def\nion{n_{\rm i}}
\def\nprot{n_{\rm p}}
\def\pelec{p_{\rm e}}
\def\pfermi{p_{\rm f}}
\def\efermi{E_{\rm f}}
\def\mue{\mu_{\rm e}}

\def\gntff{\overline{g}_{\rm ff}}

\def\eneutrino{\nu_{\rm e}}
\def\eaneutrino{\overline{\nu}_{\rm e}}
\def\mneutrino{\nu_{\mu}}
\def\maneutrino{\\overline{nu}_{\mu}}
\def\tneutrino{\nu_{\tau}}
\def\maneutrino{\\overline{nu}_{\tau}}

\def\fine{\alpha_{\rm f}}
\def\comptw{\lambda_{\rm e}}

\def\threehe{^3{\rm He}}
\def\fourhe{^4{\rm He}}
\def\sixli{^6{\rm Li}}
\def\sevenli{^7{\rm Li}}
\def\ninebe{^9{\rm Be}}
\def\tenb{^{10}{\rm B}}
\def\elevenb{^{11}{\rm B}}
\def\twelvec{^{12}{\rm C}}
\def\sixteeno{^{16}{\rm O}}

\def\fourhebyh{\fourhe/{\rm H}}
\def\threehebyh{\threehe/{\rm H}}
\def\sevenlibyh{^7{\rm Li}/{\rm H}}
\def\bsqfh{{\rm B^2FH}}

\def\mag{\unit{magnitude}}
\def\mags{\unit{magnitudes}}
\def\absm#1{M_{\rm#1}}
\def\appm#1{m_{\rm#1}}
\def\Delm#1{\Delta m_{#1}}
\def\magarcsecsqi{\unit{mag\arcsecsqi}}

\def\lsol{L_{\odot}}
\def\msol{M_{\odot}}
\def\rsol{R_{\odot}}
\def\zsol{Z_{\odot}}

\def\aunit{\unit{AU}}

\def\mear{M_{\oplus}}
\def\rear{R_{\oplus}}
\def\aear{a_{\oplus}}
\def\pear{P_{\oplus}}

\def\mjup{M_{\rm Jupiter}}
\def\ajup{a_{\rm Jupiter}}
\def\pjup{P_{\rm Jupiter}}
\def\msat{M_{\rm Saturn}}

\def\amer{a_{\rm Mercury}}
\def\pmer{P_{\rm Mercury}}

\def\aplu{a_{\rm Pluto}}
\def\pplu{P_{\rm Pluto}}

\def\meight{\bb\frac{M}{\onlyten{8}\msol}\eb}

\def\mbymsol{\bb\frac{M}{\msol}\eb}

\def\aop{\alpha_{\rm op}}
\def\ar{\alpha_{\rm R}}
\def\ax{\alpha_{\rm x}}
\def\ag{\alpha_{\gamma}}
\def\aire{\alpha_{2.2-25\micron}}
\def\asm{\alpha_{\rm sm}}
\def\onemal{1-\alpha}
\def\nuone{\nu_1}
\def\nutwo{\nu_2}
\def\nuz{\nu_0}
\def\nug{\nu_g}
\def\nuc{\nu_c}
\def\nua{\nu_a}
\def\nup{\nu_p}
\def\nux{\nu_{\rm x}}
\def\nuop{\nu_{\rm op}}
\def\nupz{\nu_{p0}}
\def\nuaop{\nu^{-\alpha_{op}}}
\def\nualpha{\nu^{-\alpha}}
\def\alnu{\alpha_{\nu}}
\def\alnua{\alpha_{\nu_a}}
\def\alnusc{\alnu^{\rm sc}}
\def\jnu{j_{\nu}}
\def\jnua{j_{\nu_a}}
\def\tstar{\tau_{*}}
\def\tnu{\tau_{\nu}}
\def\etnu{e^{\tnu}}
\def\emtnu{e^{-\tnu}}
\def\fnu{F(\nu)}
\def\fnua{F(\nua)}
\def\fr{F_{\rm R}}
\def\fop{F_{\rm op}}
\def\fopl{F_{\rm op}^l}
\def\fx{F_{\rm x}}
\def\fxl{F_{\rm x}^{l}}
\def\frlim{F_{R,lim}}
\def\mv{m_V}
\def\mb{m_B}
\def\mr{m_R}
\def\Mv{M_V}
\def\Mb{M_B}
\def\Mr{M_R}

\def\bnu{B(\nu)}
\def\blnu{B_{\nu}}
\def\inu{I(\nu)}
\def\ilnu{I_{\nu}}
\def\snu{S(\nu)}
\def\slnu{S_{\nu}}
\def\Snu{S(\nu)}
\def\unu{u(\nu)}
\def\ulnu{u_{\nu}}
\def\Snua{S(\nu_a)}
\def\nnu{n(\nu)}
\def\nlnu{n_{\nu}}
\def\nomega{n(\omega)}
\def\nlomega{n_{\omega}}
\def\taunu{\tau_{\nu}}
\def\etaunu{e^{\taunu}}
\def\emtaunu{e^{-\taunu}}
\def\exptaunu{\exp(\taunu)}
\def\expmtaunu{\exp(-\taunu)}
\def\planckinu{\frac{2h\nu^3/c^2}{\exp(h\nu/kT)-1}}
\def\planckunu{\frac{8\pi h\nu^3/c^3}{\exp(h\nu/kT)-1}}
\def\planckilambda{\frac{2hc^2/\lambda^5}{\exp(hc/\lambda kT)-1}}
\def\planckulambda{\frac{8\pi hc/\lambda^5}{\exp(hc/\lambda kT)-1}} 

\def\plisnu{\frac{2h\nu^3/c^2}{\exp(h\nu/kT)-1}}
\def\plusnu{\frac{8\pi h\nu^3/c^3}{\exp(h\nu/kT)-1}}
\def\plislambda{\frac{2hc^2/\lambda^5}{\exp(hc/\lambda kT)-1}}
\def\pluslambda{\frac{8\pi hc/\lambda^5}{\exp(hc/\lambda kT)-1}} 
\def\plnut{\frac{2h\nu^3/c^2}{\exp(h\nu/kT)-1}}
\def\rjnut{\frac{2kT\nu^2}{c^2}}

\def\Pnu{P(\nu)}
\def\lnu{L(\nu)}
\def\llnu{\log \lnu}
\def\lr{L_{\rm R}}
\def\llr{\log L_{\rm R}}
\def\lrlim{L_{R,lim}}
\def\lfive{L(5\ghz)}
\def\lop{L_{{\rm op}}}
\def\llop{\log L_{{\rm op}}}
\def\lx{L_{\rm x}}
\def\llx{\log L_{\rm x}}
\def\lxmin{L_{\rm x,min}}
\def\lxmax{L_{\rm x,max}}
\def\lmax{L_{|rm max}}
\def\lir{L_{\rm IR}}
\def\luv{L_{\rm UV}} 
\def\lrc{L_{\rm rc}}
\def\llrc{\log L_{\rm rc}}
\def\lrext{L_{\rm \rm r,ext}}
\def\llrext{\log L_{r,ext}}
\def\lre{L_{\rm re}}
\def\llre{\log L_{\rm re}}
\def\lxb{L_{\rm xb}}
\def\llxb{\log L_{\rm xb}}
\def\lxu{L_{\rm xu}}
\def\llxu{\log L_{\rm xu}}
\def\lre{L_{\rm re}}
\def\llre{\log L_{\rm re}}
\def\rt{R_{\rm t}}
\def\rx{R_{\rm x}}
\def\rxbu{R_{\rm xbu}}
\def\rtx{R_{\rm tx}}
\def\gxbp{g_{\rm x}(\beta, \psi)}
\def\grbp{g_{\rm r}(\beta, \psi)}
\def\aox{\alpha_{\rm ox}}
\def\aoxl{\alpha_{\rm ox}^l}
\def\aro{\alpha_{\rm ro}}
\def\arx{\alpha_{\rm rx}}
\def\aoxx{\alpha_{\rm oxx}}
\def\auv{\alpha_{\rm uv}}
\def\sigt{\sigma_{\rm T}}
\def\sigkn{\sigma_{\rm KN}}
\def\signu{\sigma_{\nu}}
\def\siggg{\sigma_{\gamma\gamma}}
\def\tausc{\tau_{\rm sc}}
\def\taut{\tau_{\rm T}}
\def\nsc{N_{\rm sc}}
\def\freenu{l_{\nu}}
\def\U #1{U_{\rm #1}}
\def\eps{\epsilon}
\def\epsone{\epsilon_1}
\def\epstwo{\epsilon_2}
\def\epsx{\epsilon_{\rm x}}
\def\epsxs{\eps_{\rm xs}}
\def\epss{\epsilon_{\rm s}}
\def\epsg{\epsilon_{\gamma}}
\def\epsnu{\epsilon_{\nu}}
\def\epsir{\eps_{\rm IR}}
\def\epsi{\eps_{\rm i}}
\def\epsf{\eps_{\rm f}}
\def\DOM{\Delta\Omega}
\def\omegaone{\omega_{1}}

\def\hnu{h\nu}
\def\hnubyc{\frac{h\nu}{c}}
\def\hbarom{\hbar\omega}

\def\lcapc{L_{\rm C}}
\def\lcaps{L_{\rm S}}
\def\lcapi{L{\rm i}}
\def\lcaph{L_{\rm h}}
\def\dtvar{\Delta t_{\rm var}}
\def\tbri{T_{\rm b}}
\def\tsri{T_{\rm i}}
\def\Tvar{T_{\rm var}}
\def\dmv{dm_{\rm V}}
\def\betaa{\beta_{\rm a}}
\def\vela{v_{\rm a}}
\def\te{t_{\rm e}}
\def\dte{dt_{\rm e}}
\def\gammap{\gamma_{\rm p}}
\def\gammab{\gamma_{\rm b}}
\def\gammai{\gamma{\rm i}}
\def\gammaj{\gamma{\rm j}}
\def\gammac{\gamma{\rm c}}

\def\vbyvm{\frac {V} {V_m}}
\def\vbyvmp{\frac {V'} {V_m'}}
\def\mvbyvm{\bba\vbyvm\eba}
\def\mvbyvmp{\bba\vbyvmp\eba}

\def\vbyvml{V/V_m}
\def\vbyvmpl{V'/V_m'}
\def\mvbyvml{\bba\vbyvml\eba}
\def\mvbyvmpl{\bba\vbyvmpl\eba}

\def\bperp{B_{\perp}}
\def\bpar{B_{\parallel}}
\def\bperppar{B_{\perp-\parallel}}
\def\dtvar{\Delta t_{var}}
\def\Tvar{T_{var}}
\def\dmv{dm_v}
\def\betaa{\beta_a}
\def\gammap{\gamma_{\rm p}}
\def\gammab{\gamma_{\rm b}}
\def\gammamax{\gamma_{\rm max}}
\def\scrid{{\cal D}}
\def\scril{{\cal L}}
\def\scrio{{\cal O}}
\def\scrir{{\cal R}}
\def\fbeam{F_b}
\def\fext{F_{ext}}
\def\fin{F_{\rm in}}
\def\fout{F_{\rm out}}
\def\bcospsi{\beta\cos\psi}
\def\bsinpsi{\beta\sin\psi}
\def\rnu{R_{\nu}}
\def\rnuone{R_{\nu_1}}
\def\rnutwo{R_{\nu_2}}

\def\nh{N_{\rm H}}
\def\mh{M_{\rm H}}
\def\eax{E^{-\ax}}
\def\ex{E_{\rm x}}
\def\ex{E_{\rm xs}}

\def\zmin{z_{\rm min}}
\def\zmax{z_{\rm max}}
\def\rmin{r_{\rm min}}
\def\rmax{r_{\rm max}}
\def\rsc{r_{\rm sc}}
\def\rscm{r_{\rm sc,m}}

\def\corrxz{r_{\rm x,z}}
\def\corrxop{r_{\rm x,op}}
\def\corropz{r_{\rm op,z}}
\def\corrxopz{r_{\rm x,op;z}}
\def\chisq{\chi^2}
\def\chisqr{\chi^{2}_{\nu}}

\def\philx{\Phi{(\lx)}}
\def\phixstar{\Phi_{X}^{*}}
\def\phixstaro{\Phi_{X1}^{*}}
\def\lxstar{\lx^{*}}
\def\lxfour{L_{{\rm x},44}}
\def\lxstarfour{L_{{\rm x},44}^*(0)}

\def\asca{{\em ASCA }}
\def\axaf{{\em AXAF }}
\def\ein{{\em EINSTEIN }}
\def\exosat{{\em EXOSAT }}
\def\ginga{{\em GINGA }}
\def\rosat{{\em ROSAT }}
\def\compton{{\em COMPTON }}
\def\granat{{\em GRANAT }}
\def\iras{{\em IRAS }}
\def\iue{{\em IUE }}

\def\seyone{Seyfert\,1 }
\def\seytwo{Seyfert\,2 }
\def\bllac{BL Lac }
\def\bllacs{BL Lacs }
\def\rbl{{\rm RBL }}
\def\xbl{{\rm XBL }}
\def\ipc{{\rm IPC }}
\def\pspc{{\rm PSPC }}

\def\cfour{{\rm C\,IV }}
\def\cthreesf{{\rm C\,III] }}
\def\fetwo{{\rm Fe\,II }}
\def\halpha{{\rm H\,\alpha }}
\def\hbeta{{\rm H\,\beta }}
\def\hgamma{{\rm H\,\gamma}}
\def\hi{{\rm H\,I }}
\def\lyal{{\rm Ly\,\alpha }}
\def\mgtwo{{\rm Mg\,II }}
\def\ntwo{{\rm [N\,II]}}
\def\nfive{{\rm N\,V }}
\def\otwo{{\rm [O\,II] }}
\def\othree{{\rm [O\,III] }}
\def\kalpha{K_{\alpha} }
\def\ykz{Y^{K}_{Z} }

\def\const{{\rm constant}}
\def\gray{\gamma-{\rm ray}}
\def\grays{\gamma-{\rm rays}}
\def\maxtau{\tausc(1+\tausc)}
\def\maxtnu{\tnu(1+\tnu)}
\def\dm{{\rm DM}}
\def\dmsinmb{\dm\sin\mod{b}}
\def\dmsinb{\dm\sin b}
\def\sbyn{\frac{S}{N}}

\def\noi{\noindent}
\def\vf{\vspace{0.5cm}}
\def\vff{\vspace{1cm}}
\def\pbr{\pagebreak}
\def\pnew{\newpage}
\def\pemp{\pagestyle{empty}}
\def\tpemp{\thispagestyle{empty}}

\newcommand{\onlytitle}[1]
{\title{#1}
\author{ }
\date{ }
\maketitle\tpemp }

\newcommand{\onlyfig}[2]
{\begin{center}\epsfig{figure=#1.ps, width=#2in}\end{center}}

\newcommand{\onlyfige}[2]
{\begin{center}\epsfig{figure=#1.eps, width=#2in}\end{center}}

\def\vbyc{\frac{v}{c}}
\def\vbycsq{\frac{v^2}{c^2}}
\def\omvbycsq{\sqrt{1-\vbycsq}}

\def\gmr{\frac{GM}{r}}
\def\gmcsr{\frac{GM}{c^2r}}
\def\gmcsrl{GM/c^2r}
\def\tgmcsr{\frac{2GM}{c^2r}}
\def\schwarz{r_{\rm S}}
\def\tschwarz{t_{\rm S}}
\def\rergo{r_{\rm E}}
\def\rplus{r_{+}}
\def\mbh{M_{\rm BH}}
\def\mbyl{M/L}

\def\lh{l_{\rm h}}
\def\li{l_{\rm i}}
\def\ls{l_{\rm s}}

\def\omk{\Omega_{\rm K}(r)}
\def\velr{v_{r}}
\def\velp{v_{\phi}}
\def\velz{v_{z}}
\def\sigrt{\Sigma(r,t)}
\def\mach{{\cal M}}
\def\scrir{{\cal R}}
\def\tin{t_{\rm in}}
\def\tl{t_{\rm L}}
\def\ldisk{L_{\rm disk}}
\def\temps{T_{\rm s}}
\def\tempc{T_{\rm c}}
\def\tempb{T_{\rm B}}
\def\bnu{B_{\nu}}
\def\bnuts{B_{\nu}(\temps)}
\def\bnutsr{B_{\nu}(\temps(r))}

\def\rout{r_{\rm out}}
\def\tvis{t_{\rm vis}}
\def\pgas{p_{\rm g}}
\def\prad{p_{\rm r}}
\def\csou{c_{\rm s}}
\def\rstar{r_{*}}
\def\ledd{L_{\rm Edd}}
\def\mdedd{\dot{M}_{\rm Edd}}
\def\tedd{t_{\rm Edd}}
\def\tempedd{T_{\rm Edd}}
\def\bedd{B_{\rm Edd}}
\def\mdot{{\dot M}}
\def\geff{\vec{g}_{\rm eff}}
\def\phir{\Phi_{\rm rot}}
\def\phie{\Phi_{\rm eff}}
\def\zsr{z_{\rm s}(r)}
\def\udzero{u_0}
\def\udphi{u_{\phi}}
\def\uuzero{u^0}
\def\uuphi{u^{\phi}}

\def\isnu{I_{\nu}}
\def\islambda{I_{\lambda}}
\def\fsnu{F_{\nu}}
\def\fslambda{F_{\lambda}}

\def\plisnu{\frac{2h\nu^3/c^2}{\exp(h\nu/kT)-1}}
\def\plusnu{\frac{8\pi h\nu^3/c^3}{\exp(h\nu/kT)-1}}
\def\plislambda{\frac{2hc^2/\lambda^5}{\exp(hc/\lambda kT)-1}}
\def\pluslambda{\frac{8\pi hc/\lambda^5}{\exp(hc/\lambda kT)-1}} 

\def\isnurj{\isnu^{\rm RJ}}
\def\isnuw{\isnu^{\rm W}}

\def\cv{C_V}
\def\cp{C_p}
\def\dqdtv{\bb\pderi{Q}{T}\eb_V}
\def\dqdtp{\bb\pderi{Q}{T}\eb_p}

\def\CH{{\it CH}}

\def\hub{H_0}
\def\hubi{H^{-1}_0}
\def\hubble#1{H_0={#1}\unit{km\,sec^{-1}\,Mpc^{-1}}}
\def\hubbleunit{\unit{km\,sec^{-1}\,Mpc^{-1}}}
\def\hubhund{h_{100}}
\def\hubhundcu{\hubhund^3}
\def\hubhundcui{\hubhund^{-3}}
\def\hubhundi{\hubhund^{-1}}
\def\hubhundis{\hubhund^{-2}}
\def\cuponh{\bb \frac {c} {H_0} \eb}

\def\qz{q_0}
\def\qzsq{\qz^2}
\def\fqz{\bbcu\reci{\qzsq}\bbsq \qz z + (q_0-1)(\sqrt{1+2q_0z}-1)
\ebsq\ebcu}
\def\fqzno{\qz z + (q_0-1)(\sqrt{1+2q_0z}-1)}
\def\omb{\Omega_{{\rm b}}}
\def\omhi{\Omega_{{\rm H\,I}}}
\def\omlls{\Omega_{{\rm LLS}}}
\def\omlb{\Omega_{{\rm lb}}}
\def\omigm{\Omega_{{\rm IGM}}}
\def\omdlyal{\Omega_{{\rm DLy\,\alpha}}}
\def\zpr{z^{\prime}}
\def\zabs{z_{{\rm abs}}}
\def\zem{z_{{\rm em}}}
\def\zlim{z_{\rm lim}}
\def\zlimg{z_{\rm lim,G}}
\def\zlimgq{z_{\rm lim,GQ}}
\def\dlum{D_{\rm L}}
\def\dlumsq{D^2_{\rm L}}
\def\dmet{d_{\rm m}}
\def\dang{d_{\rm A}}

\def\lamo#1{\lambda_{\rm o_{#1}}}
\def\lame#1{\lambda_{\rm e_{#1}}}

\def\re{r_{\rm e}}
\def\rbyre{\frac{r}{\re}}
\def\rbyredev{\bb\rbyre\eb^{1/4}}
\def\rbyren{\bb\rbyre\eb^{1/n}}
\def\rbyrel{r/\re}
\def\rbyredevl{(\rbyrel)^{1/4}}
\def\rbyrenl{(\rbyrel)^{1/n}}
\def\rebyrd{\frac{\re}{\rd}}
\def\rebyrdl{\re/\rd}

\def\izero{I(0)}
\def\ire{I(\re)}
\def\ir{I(r)}
\def\icr{I_{\rm c}(r)}

\def\muzero{\mu(0)}
\def\mure{\mu(\re)}
\def\mur{\mu(r)}

\def\ibzero{I_{\rm b}(0)}
\def\ibre{I_{\rm b}(\re)}
\def\mibre{\mean{\ibre}}
\def\iblre{I_{\rm b}(<\re)}
\def\ibr{I_{\rm b}(r)}
\def\iblr{I_{\rm b}(<r)}
\def\ibtot{I_{\rm b,tot}}
\def\ibxy{I_{\rm b}(x,y)}
\def\isky{I_{\rm sky}}

\def\mubzero{\mu_{\rm b}(0)}
\def\mubre{\mu_{\rm b}(<\re)}
\def\mmubre{\mean{\mubre}}
\def\mubr{\mu_{\rm b}(r)}
\def\mulim{\mu_{\rm lim}}

\def\mudzero{\mu_{\rm d}(0)}
\def\mugzero{\mu_{\rm G}(0)}

\def\rd{r_{\rm d}}
\def\rs{r_{\rm s}}
\def\rbyrd{\frac{r}{\rd}}
\def\rbyrdl{r/\rd}
\def\idzero{I_{\rm d}(0)}
\def\idr{I_{\rm d}(r)}
\def\idlr{I_{\rm d}(<r)}
\def\idtot{I_{\rm d,tot}}
\def\idxy{I_{\rm d}(x,y)}

\def\imax{I_{\rm max}}
\def\imin{I_{\rm min}}

\def\dev{\onlyten{-3.33\rbyredevl}}
\def\devn{\onlyten{-\cn\rbyren}}
\def\exprbyrd{e^{-(\rbyrdl)}}
\def\cn{c_n}

\def\dbyb{\frac{D}{B}}
\def\bbyd{\frac{B}{D}}
\def\dbybl{D/B}
\def\bbydl{B/D}

\def\absgal{M_{\rm G}}
\def\absbul{M_{\rm B}}
\def\absdisk{M_{\rm d}}
\def\absq{M_{\rm Q}}

\def\lstar{L_{*}}

\def\mbyl{\bb\frac{M}{L}\eb}

\def\labchap #1{\label{chap:#1}}
\def\labequn #1{\label{eq:#1}}
\def\labfig #1{\label{fig:#1}}
\def\labsecn #1{\label{sec:#1}}
\def\labsubsecn #1{\label{subsecn:#1}}
\def\labsubsubsecn #1{\label{subsubsecn:#1}}
\def\labtablem #1{\label{tab:#1}}
\def\labapp #1{\label{app:#1}}
\def\labboxl #1{\label{box:#1}}

\def\boxl #1{box~\ref{box:#1}}
\def\chap #1{Chapter~\ref{chap:#1}}
\def\equn #1{Equation~\ref{eq:#1}}
\def\fig #1{Figure~\ref{fig:#1}}
\def\relation #1{relation~\ref{eq:#1}}
\def\secn #1{Section~\ref{sec:#1}}
\def\subsecn #1{Section~\ref{subsecn:#1}}
\def\subsubsecn #1{Section~\ref{subsubsecn:#1}}
\def\app#1{Appendix~\ref{app:#1}}
\def\tablem #1{Table~\ref{tab:#1}}

\def\dequn#1#2{Equations~{\ref{eq:#1}}~and~{\ref{eq:#2}}}
\def\tequn#1#2#3{Equations~{\ref{eq:#1}},~{\ref{eq:#2}} and ~{\ref{eq:#3}}}
\def\mequn#1#2{Equations~{\ref{eq:#1}}~to~{\ref{eq:#2}}}
\def\dfig#1#2{Figures~{\ref{fig:#1}}~and~{\ref{fig:#2}}}
\def\mfig#1#2{Figures~{\ref{fig:#1}}~to~{\ref{fig:#2}}}
\def\dtablem#1#2{Tables~{\ref{tab:#1}}~and~{\ref{tab:#2}}}
\def\dsubsecn#1#2{Subsections~{\ref{subsecn:#1}}~and~{\ref{subsecn:#2}}}

\def\tobs#1{{\it The Observatory}{\bf #1}}

\def\cfa#1{{\it Center for Astrophysics Preprint }{\bf #1}}
\def\asp{Astron. Soc. of the Pacific, San Francisco.}
\def\clar{Clarendon Press, Oxford.} 
\def\cup{Cambridge University Press, Cambridge.}
\def\free{W. H. Freeman, New York.}
\def\klu{Kluwer, Dordrecht.}
\def\nrao{NRAO, Greenbank.}
\def\plenumn{PLenum Press, New York.}
\def\prince{Princeton University Press, Princeston}
\def\ridel{Ridel, Dordrecht.}
\def\springerb{Springer-Verlag, Berlin.}
\def\springern{Springer-Verlag, New York.}
\def\wiley{Wiley, New York.}
\def\yale{Yale University Press, New Haven.}

\def\noin{\noindent}

\def\sqr#1#2{{\vcenter{\vbox{\hrule height.#2pt
  \hbox {\vrule width.#2pt height#1pt \kern#1pt
  \vrule width.#2pt}
  \hrule height.#2pt}}}}

\def\square{\mathchoice\sqr68\sqr68\sqr{2.1}3\sqr{1.5}3}

\def\CH{{\it CH }}

\newcommand{\myruleb}[1]{\vspace{0.5cm}\noindent\rule{#1}{0.05in}}
\newcommand{\myrulee}[1]{\newline\noindent\rule{#1}{0.05in\vspace{0.5cm}}}

\newenvironment{mybox}[3]
{\begin{minipage}{4.5in}
\setlength{\parindent}{0.5cm}
\myruleb{4.5in}
\label{box:#1}
\begin{center}\fbox{\textbf{#2}}\end{center}
{\begin{small}#3\end{small}}
\myrulee{4.5in}
\end{minipage}}

\newenvironment{myboxp}[4]
{\begin{minipage}{4.5in}
\setlength{\parindent}{0.5cm}
\myruleb{4.5in}
\label{box:#1}
\begin{center}\fbox{\textbf{#2}}\end{center}
\begin{center}\epsfig{figure=#3.ps,width=2in}\end{center}
{\begin{small}#4\end{small}}
\myrulee{4.5in}
\end{minipage}}

\newcommand{\myfig}[2]
{\begin{figure}[ht]
\begin{center}
\label{fig:#1}
\epsfig{figure=#1.ps,width=5in}
\caption{#2}
\end{center}
\end{figure}}

\newcommand{\myplotone}[2]
{\begin{figure}[ht]
\begin{center}
\label{fig:#1}
\plotone{#1.ps}
\caption{#2}
\end{center}
\end{figure}}

\newcommand{\myequn}[2]
{\begin{equation}
{#2}
\labequn{#1}
\end{equation}}

\newcommand{\myequna}[2]
{\begin{eqnarray}
{#2}
\labequn{#1}
\end{eqnarray}}

\newenvironment{headnoff}[1]
{\pnew\bHuge\bcenter{\bf\underline{#1}}\ecenter\eHuge}

\newcommand{\state}[1]{\noindent{#1}\vf}

\newcommand{\mnote}[1]{{{\small\it\marginpar{#1}}}}

\def\dev{r^{1/4}~}
\def\ser{r^{1/n}~}
\def\sbe{{\mu_{\rm b}(\re)}}            
\def\msbe{\mmubre}                      
\def\mie{\mean{I_{\rm b}(<\re)}}        
\def\kmps{\unit{km \, s^{-1}}}
\def\magper{\unit{mag\, arcsec^{-2}}}
\def\ul{\underline}

\def\sqb #1{\left[#1\right]}
\def\sib #1{\left(#1\right)}
\def\cub #1{\left{#1\right}}
\newcommand{\plotn}[2]   
{\begin{figure*}
\centering
\includegraphics[width=\textwidth]{#1.eps}
\caption{#2}
\label{#1}%
\end{figure*}}

\begin {document}
\title{Photometric Scaling Relations for Bulges of Galaxies}
\author{C. D. Ravikumar,\inst{1}
Sudhanshu Barway, \inst{2}
Ajit Kembhavi,\inst{3}
Bahram Mobasher,\inst{4}
\and
V. C. Kuriakose\inst{5}}

\offprints{C. D. Ravikumar}

\institute{GEPI, Observatoire de Paris-Meudon, 92195 Meudon, France
\\ \email{Chazhiyat.Ravikumar@obspm.fr}
\and
School of Studies in Physics, Pt. Ravishankar Shukla
University, Raipur, India. \\ \email{sudhan@iucaa.ernet.in} 
\and
Inter University Centre for Astronomy and Astrophysics, 
Post Bag 4, Ganeshkhind, Pune 411007, India. \\ \email{akk@iucaa.ernet.in}
\and
Space Telescope Science Institute, 3700 San Martin
Drive, Baltimore MD 21218, USA. \email{mobasher@stsci.edu}
\and
Department of Physics, Cochin University of Science 
and Technology, Kochi 682022, India. \\ \email{vck@cusat.ac.in}
}

\date{Received xxxx; accepted xxxx}

\abstract
{We study the photometric parameters of the bulges of galaxies of
different Hubble types including
ellipticals, lenticulars, early and late type spirals and early type 
dwarf galaxies. Analyzing the distributions of various photometric 
parameters, and two- and 
three-dimensional correlations between them, we  find that there is 
a difference in the correlations exhibited by bright ($M_{\rm K} < -22$) and
faint bulges, irrespective of their  Hubble type. Importantly, the bright 
bulges, which include 
typically E/S0 galaxies and bulges of early type spirals, are tightly
distributed around a  common 
photometric plane (PP), while their fainter counter parts, mainly 
bulges of late type spirals and dwarf galaxies 
show significant deviation from the planar distribution. We show that
the specific entropy, determined from the bulge structural parameters,
systematically  increases as we move from late to early Hubble
types. We interpret this as 
evidence for hierarchical merging and passive
evolution scenarios for bright and faint bulges respectively.

\keywords{galaxies: clusters: individual.. - galaxies: bulges - galaxies: fundamental parameters - galaxies: structure}
}
\authorrunning{Ravikumar \etal}
\titlerunning{Photometric Scaling Relations for Galaxies}
\maketitle

\section{Introduction}
\labsecn{intro}
Galaxies show a wide variety of morphologies. Right from the work of Hubble 
(1936) attempts have been  made to classify them according to their apparent shape, 
structure, elongation \etc The wide ranges seen  in their 
properties like luminosity and color, and in parameters defining their large
scale structure, indicate that galaxies are formed through different mechanisms and
are in a state of constant evolution.  A detailed 
analysis of the distribution of such properties, and the correlations between them, 
helps in understanding the systematics involved 
in the structure, formation and  evolution of galaxies. 

The ratio of bulge to disk luminosity in galaxies decreases along the Hubble
sequence. Elliptical galaxies (Es), which are wholly
dominated by their bulges, show considerable  homogeneity, mainly characterized by 
the existence of the fundamental plane (FP), a tight correlation between
central velocity dispersion ($\sigma$)- effective radius ($\re$)-
effective mean surface brightness ($\msbe$)- (\eg, Dressler
\etal 1987; Djorgovski \& Davis 1987) with little intrinsic 
scatter.  The homogeneity extends to many other correlations 
as well, like the color-magnitude (CM, Bower \etal 1992; Andreon 2003), 
the Faber-Jackson (1976) and Kormendy (1985) relations. Lenticulars (S0s) 
show striking similarities with Es in their abundance near cluster centers.
The properties of the bulges of lenticulars follow closely the relations 
obeyed by Es, while their disks show structural 
similarities with the disks  of spiral galaxies (Mathieu \etal 2002), sometimes
 with larger scatter (Hinz \etal 2003) towards their faint end. Spiral 
galaxies have more complex structures, with the disk having spiral arms whose 
characteristic too change along the Hubble sequence.  In this paper, 
we intend to study the photometric properties of the 
bulges of different Hubble types for galaxies.

A systematic and quantitative understanding of the morphological structure in 
galaxies is obtained by analyzing their  surface brightness (SB) distribution  
to separate their bulge and disk properties. 
In the simplest approach, this is 
achieved by parameterizing the radial structure of the SB profile in terms of 
empirical functions with a small number of free parameters.    
The SB profile of discs is satisfactorily  fitted 
by an exponential function characterized by a central surface brightness and a 
disk scale length (Freeman 1970; also see 
Grosb{\o}l 1985; Courteau 1996). The situation with bulges is somewhat 
more complicated.  The bulge dominated surface brightness profiles of
ellipticals were found to be reasonably well fitted by the two parameter 
de Vaucouleurs' $\dev$ law (de Vaucouleurs, 1959).  
However, in recent years better quality CCD data 
have shown that the generalization of $\dev$ to a $\ser$ 
law, first proposed by S\'ersic (1968), represents the  SB profiles better. 
The $\ser$ law also provides
satisfactory fits to the bulges of lenticular and spiral galaxies. 
The significant correlations that the S\'ersic index $n$ shares with 
other photometric parameters  like luminosity and central surface 
brightness (Young \& Currie 1994; Andredakis, Peletier, \& Balcells 1995; 
Binggeli \& Jerjen 1998; Khosroshahi \etal 2000a,b; Trujillo, Graham 
\& Caon 2001, Trujillo \etal 2002, Gerbal \etal 1997, Lima-Neto
\etal 1999,  M\'arquez \etal 2000;2001), with spectroscopic parameters like
central velocity dispersion  
$\sigma$ (\eg Graham \& Colless 1997) and [Mg/Fe] abundance ratios (Vazdekis, 
Trujillo,\& Yamada 2004) and also with the mass of the galaxy's (central) 
super-massive black hole (Graham \etal 2001)  suggest that this
generalization provides important information regarding the physical processes 
involved in the evolution of the bulges.   
     
An attempt to understand and interpret the observed correlations
  from a theoretical point of view has been made by Gerbal \etal (1997), 
  Lima-Neto \etal (1999), and  M\'arquez \etal (2001).  These authors
  consider  elliptical galaxies to be in a state of quasi-equilibrium
  characterized by a local maximum in the specific entropy of the system.
  Studies of simulations of successive
  mergers of elliptical galaxies showed a systematic increase in the
  specific entropy with the merging history (M\'arquez \etal 2000),
  suggesting that the specific entropy can be used as a potential tool in
  understanding the evolution of galaxies.

        In this paper, we will consider the quantitative decomposition of
galaxies of different Hubble types into the bulge and disk components, and 
consider in detail various relations between the bulge properties.  We
will differ the
discussion of disk properties to a later work. The paper is arranged as 
follows.  In \S 2 we introduce the sample used 
in this analysis. Section 3 deals with the determination of the 
structural parameters of early type galaxies in $K$ band for 
two rich clusters Abell 
2199 and Abell 2634 using the full two dimensional bulge-disc 
decomposition algorithm, \emph{fitgal}. We then move on to describe 
the various photometric properties in \S 4 and show that the brighter 
($M_{\rm K} < -22$) and fainter bulges, irrespective of their Hubble type, 
fill distinct regions in their parametric correlations.  Assuming 
gravo-thermal 
properties to the bulges, we then estimate their specific entropies 
in \S 5 and attempt to provide a robust evolution scenario of 
galaxies. Finally, in \S 6 we conclude with 
a discussion of our results and their implications.

\section {The Sample}
\labsecn{sample}

The objects of our study consist of the bulges of galaxies of different
types. For this purpose we have compiled samples of  cluster ellipticals,
lenticulars, early and late type spirals, and early type  dwarf galaxies. 
All the galaxies in our sample, except the dwarf galaxies, have been observed 
in the near infra red, the advantages of this band being its sensitivity 
towards the old population of stars and the capability to 
penetrate the dust.     
A brief description of the sample is given below.
\begin{itemize}

\item A set of 34 elliptical galaxies 
in two nearby clusters, Abell 2199 (20 galaxies) and A2634 (14 galaxies) 
at redshift $z=0.031$ and $0.030$ respectively. The sample contains
elliptical galaxies that are 
spectroscopically confirmed 
members of these clusters with reliable velocity dispersion and optical
photometry from a study by 
Lucey \etal (1997).   These galaxies were chosen to have optical half-light 
radius less than 1 arcmin, to fit in the 
field of view of the Infra-red Array Detector (IRCAM3) on the United
Kingdom Infra-red Observatory (UKIRT).  The bulge disk decomposition
algorithm we used for determining the photometric structural
parameters requires that the surface profile be fitted by a smooth
bulge + disk model. Hence galaxies with complicated morphologies, 
like strongly interacting galaxies,  were removed from the
original sample. However such 
galaxies constitute only a small fraction of the sample, and 
hence their removal posed no threat to the statistical
completeness of the sample.  It may be noted that this final criterion
is applicable to all the following samples. 
\item A set of 42 ellipticals from the Coma cluster analyzed by
Khosroshahi \etal (2000b). These galaxies were taken from a previous
study (Mobasher \etal 1999) which originally contained 48 ellipticals
applying the same selection criteria for our Abell cluster ellipticals
from the spectroscopic and optical photometric catalogue by Lucey
\etal (1991).
\item A sample of 37 confirmed field lenticular galaxies observed in
  the $K'$ band from the \emph{Observatorio Astronomico Nacional} 2.1
  m telescope at San Pedro Martir, Mexico, by Barway \etal (2005).
  These galaxies from the Uppsala General Catalogue (UGC) have $B <
  14$, angular diameter $D_{25} < 3$ arcmin and declination
  $5^{\circ}< \delta < 64^{\circ} $. The galaxies in our sample were
  chosen in an unbiased fashion from the larger complete sample with
  the above mentioned selection criteria, after removing possible
  mis-classifications by analyzing their $K$ band images. 
\item A set of 26 bulges of mostly \emph{early-type} spirals from a 
magnitude and size limited complete sample from the UGC 
constructed by Balcells \& Peletier (1994) from observations 
in the $K$ band. The sample is complete with right ascension between
$13^h$ and $24^h$, declination above $-2^{\circ}$, spirals  earlier than
Sc, $B < 14$, major axis diameter larger than 2 arcmin, absolute
galactic latitude larger than $20^{\circ}$ and axis ratio in $B$ larger
than 1.56. The final criterion makes the sample biased towards being 
highly inclined (inclination angle 
greater than $50^{\circ}$). Further, the smoothness of
the surface profile was guaranteed by avoiding 
galaxies that are irregular, strongly interacting, barred or with very
strong dust lines right up to the center of the galaxy.  
We have taken the bulge and disk parameters for 
this sample from the study by Khosroshahi \etal (2000a), who used
the same programs and techniques in their analysis as we do.  
\item Bulges of 40 bright ($B_{\rm T} < 12$) spiral galaxies, mostly  
\emph{late-type}, observed in $K$ band and analyzed by M\"ollenhoff 
\& Heidt (2001). The sample consists of galaxies with $B < 12$ of
Hubble type Sa to Sc, without strong bars, selected from the Revised Shapley
Ames Catalog (Sandage \& Tammann 1981). 
\item Early type dwarf galaxies from a study by Binggeli \& Jerjen 
(1998), which contains 128 highly resolved dE and dS0 profiles that 
are well described by the S\'ersic law.  These  galaxies were selected 
from the $B$ band photometry of a complete sample with $B < 18$ by 
Binggeli \& Cameron (1993). We have used the whole sample to represent early
type dwarfs, as the dEs and dSOs do not show any distinguishable
differences in their photometric correlations as shown in Binggeli \& Jerjen 
(1998), Also, we have converted the $B$ band 
photometry to that in the $K$ band assuming an average color 
$B-K = 3.2$. Given the fact that most smooth dwarf galaxies contain
homogeneous single component population, the application of constant
color is justifiable. The main motivation to  include the dwarf galaxies 
in this study is to address their properties in relation to their 
counter parts amongst other Hubble types. 
\end{itemize}

Further details on the selection and data reduction can be had 
from the references mentioned above.  We assume that Hubble's constant
and the cosmological
deceleration parameter are given by 
$H_{\rm 0} = 50\,\kmps\, \unit{Mpc^{-1}}$  and \mbox{$q_0 = 0.5$}
respectively.

\section{Morphological Parameters}
\labsecn{morphology}
Galaxies classified as elliptical in the classical catalogues
generally have smooth elliptical isophotes, with little evidence of 
morphological peculiarities, dust \etc seen in survey plates.  Later 
observations of such galaxies with CCD detectors have revealed a host 
of features, but these
typically are faint enough to still make it possible to use simple
laws to describe the large scale light distribution.  The most widely
used relation to describe the surface brightness distribution in
elliptical galaxies at optical and near infra-red wavelengths, $\ibr$,
is de Vaucouleurs law.
This law, with its two free parameters, the central surface 
brightness $\ibzero$ and the half-light radius $\re$ (defined as the 
radius of the central region of the galaxy containing half its total 
light),
provides a surprisingly good approximation to the
observed surface brightness distribution of many elliptical galaxies.  
But with high sensitivity CCD observations becoming available, deviations 
from the simple $\dev$ law have been noticed, and a three parameter 
generalization, first proposed by S\'ersic (1968),
has been widely used (see Khosroshahi \etal 2000a for a discussion
and references).  The S\'ersic law is given by
\myequn{bint}{\ibr = \ibzero e^{-2.303b_n{(r/\re)^{1/n}}},}
where $n$ is a parameter to be determined from  observations 
for each galaxy. The constant $b_n$, which depends on $n$,
is chosen so that $\re$ remains the half-light radius.  
For a given $n$, $b_n$ can be obtained as a root of the equation
\myequn{gamma}{P(2n,2.303b_n) = 0.5,  \nonumber}
where $P$($a$,$x$) is the incomplete Gamma function (see \eg Press
\etal 1992).

The surface brightness profile of the disk is well approximated to be 
an exponential (Freeman 1970), 
\myequn{dint}{\idr = \idzero e^{(-r/\rd)},}
where $\idzero$ and $\rd$ are the central disk intensity and the disk 
scale radius respectively.  
For extracting the bulge and disk parameters of the Abell
  cluster ellipticals, we have used the  
algorithm \emph{fitgal}, a full two dimensional routine, involving
$\chi_{\nu}^2$ minimization to determine the quality of the fit
obtained. This algorithm is described in Wadadekar, Robbason \&
Kembhavi (1999). 
For the Coma ellipticals and bulges of early type spirals in our sample,
the photometric parameters have been obtained using the program
\emph{fitgal}
by Khosroshahi \etal (2000b and 2000a respectively).  For the lenticulars, the
image analysis has been described by Barway \etal (2005), and
the photometric parameters have been obtained again using  \emph{fitgal} 
(Barway \etal to be submitted).  For the late type bulges and the 
dwarf galaxies 
we use results from M\"ollenhoff \& Heidt (2001) and Binggeli \& Cameron (1993)
respectively.  Even though different procedures for extracting
structural parameters can introduce some systematics in the derived
parameters, we anticipate the effect to be negligible, 
considering the fact that
all the sample galaxies were selected on the basis of the smoothness in
their surface brightness profiles.  We would like to note here that 
the definition
of morphological parameters used in some of the papers cited above can
be different. In such cases we
have made the necessary transformations to convert all parameters to the 
form used by us.  

\section{Distributions and Correlations}
\labsecn{correlations}
We consider in this section the distribution of various parameters for 
different galaxy types and correlations between the parameters. Any 
observed correlations can provide constraints on galaxy formation and 
evolution models. Also, as shown by the example of the FP and
  the  Kormendy  relation (Kormendy 1977) and the relationship between
  $n$ and bulge scale radius (Young \& Currie 1994, Lima Neto \etal
  1999), the observed correlations with small scatter can provide useful
  distance indicators to galaxies. 

\subsection{Distributions}
\plotn{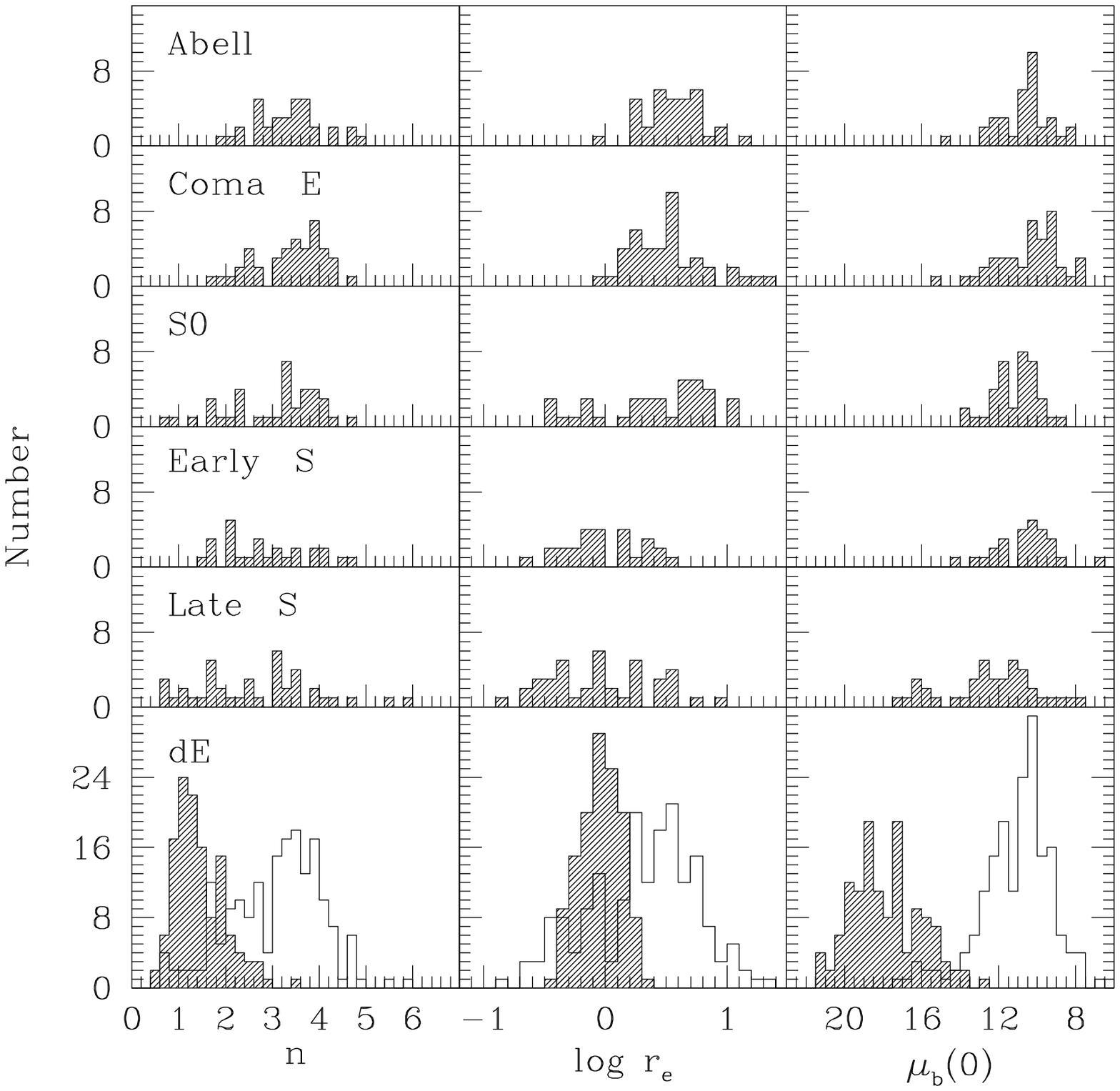}{Distribution of S\'ersic index ($n$), half-light radius 
($\log \re$ in \unit{kpc}) and unconvolved central surface brightness 
($\mubzero$ in $\magper$). The bottom panel shows histograms for dwarf
ellipticals (shaded) and the rest of the sample taken together
(open).} 
        In Fig.~\ref{sphist} we show the histogram of S\'ersic bulge parameters, 
$n$, $\log \re$, and $\mubzero$, for the different sets of galaxies in our 
sample. Clearly it can be seen that dwarf galaxies (dEs) form a different 
distribution compared to all other types. The S\'ersic index for dEs has 
a conspicuous peak at $n\sim1$, and there are very few galaxies with $n\ge2$. 
For the ellipticals $n$ peaks around $n\sim4$, with a broad spread to lower 
values (this is the reason why de Vaucouleurs' law with $n=4$ provides a 
reasonable fit to many Es). As we move along the Hubble sequence, the 
average $n$ values becomes smaller, but the spread remains large with no 
evident peak.
The distribution of $\mubzero$ for 
bulges of early type spirals show a close resemblance with  Es and bright 
S0s, while the bulges of late type spirals overlap 
with the bright end of dwarf galaxies and faint end of other early type 
bulges. The effective radius decreases 
on average along the Hubble sequence. However, contrary to the 
$\mubzero$ distribution where the dwarfs have a different distribution from 
the other types, in the case of $\re$, there is a significant overlap of the 
dEs with the other galaxies. Some bulges of S0s and spirals show smaller 
$\re$ than a majority of dEs. These distributions show that there is no 
apparent difference between the Abell and Coma cluster ellipticals in
our sample.  
Application  of the Kolmogorov-Smirnov test confirms  that  distributions 
for the two sets of cluster ellipticals are not significantly different, and 
therefore in our subsequent discussions we will be treating all the
cluster ellipticals in our sample as a single group.

        The effect of the distribution of the three basic 
parameters shown in Fig.~\ref{sphist}, on the bulge luminosity 
is such that it decreases systematically along the 
Hubble sequence. In Fig.~\ref{sphl} we show the histogram for absolute 
magnitude for the different types.
\plotn{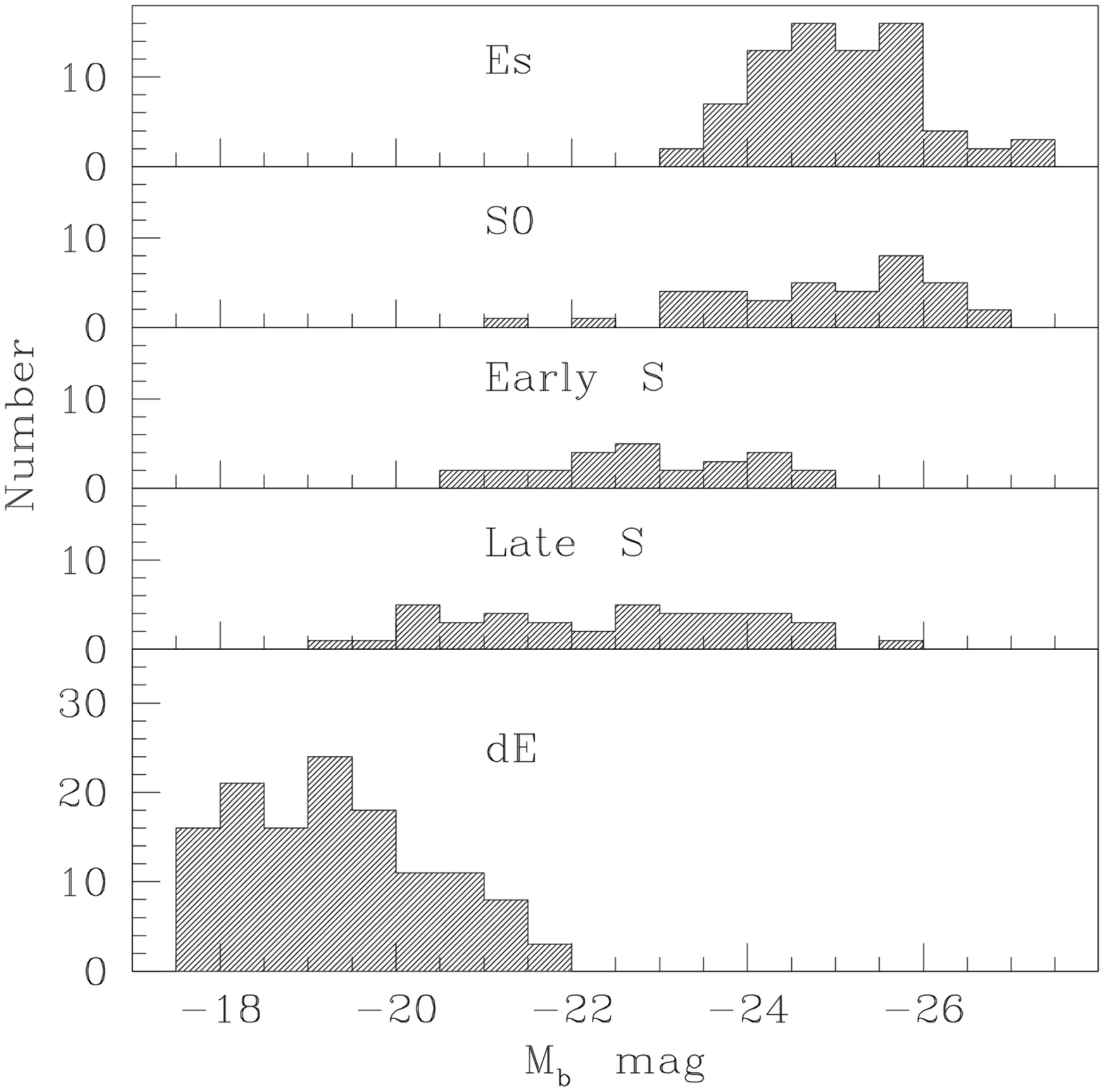}{Histogram of absolute ($K$) magnitudes for bulges. There is a 
systematic increase in the luminosity towards early type galaxies 
suggesting the dominance of bulges there.}
We have shown in Fig.~\ref{cuhist} the distribution of the three basic parameters
and the luminosity for all the bulges in our sample taken together. 
It clearly shows the division of the whole sample into 
two distinct groups. However,  there is a significant overlap of all
the bulge parameters and galaxy types in the two groups. 
This kind of a dichotomy 
in the structural parameters and their correlations has been reported 
before in the literature. Kormendy (1985) observed from his study 
on `ellipsoidal stellar systems' that bright Es show remarkable 
discontinuity in their parameter correlations from those for dwarf 
spheroidals (see also Capaccioli, Caon, \& D'Onofrio 1992).  This
has led Kormendy \& Bender (1996) to suggest a 
modifications to the Hubble classification based on the 
apparent elongation and its association with rotation. 
The dichotomy between the properties of bulges is not limited to
ellipticals. van den Bergh (1994), analyzing a sample of S0s, came to 
the conclusion that faint S0s ($M_{\rm B} > -19.5$) are statistically 
more likely to be prolate than their brighter counterparts.  
Recent studies suggest that the bulges of early and late type 
spirals seems to 
follow distinct distributions in their elongation (Fathi \& Peletier 
2003) as well as in luminosities (the luminosity distribution is
of course affected by the role that bulges play in the Hubble 
classification scheme). The dichotomy shown in 
Fig.~\ref{cuhist} does not seem to be resulting from the uneven sample size 
for the different Hubble types, since there are significant differences in 
the population of galaxies in the two groups for different parameters. 
The important observation 
we make here is that the dichotomy does not depend on the Hubble type, 
it rather seems to be a more fundamental property of the bulges that 
might be closely related to their formation and evolution processes. 
\plotn{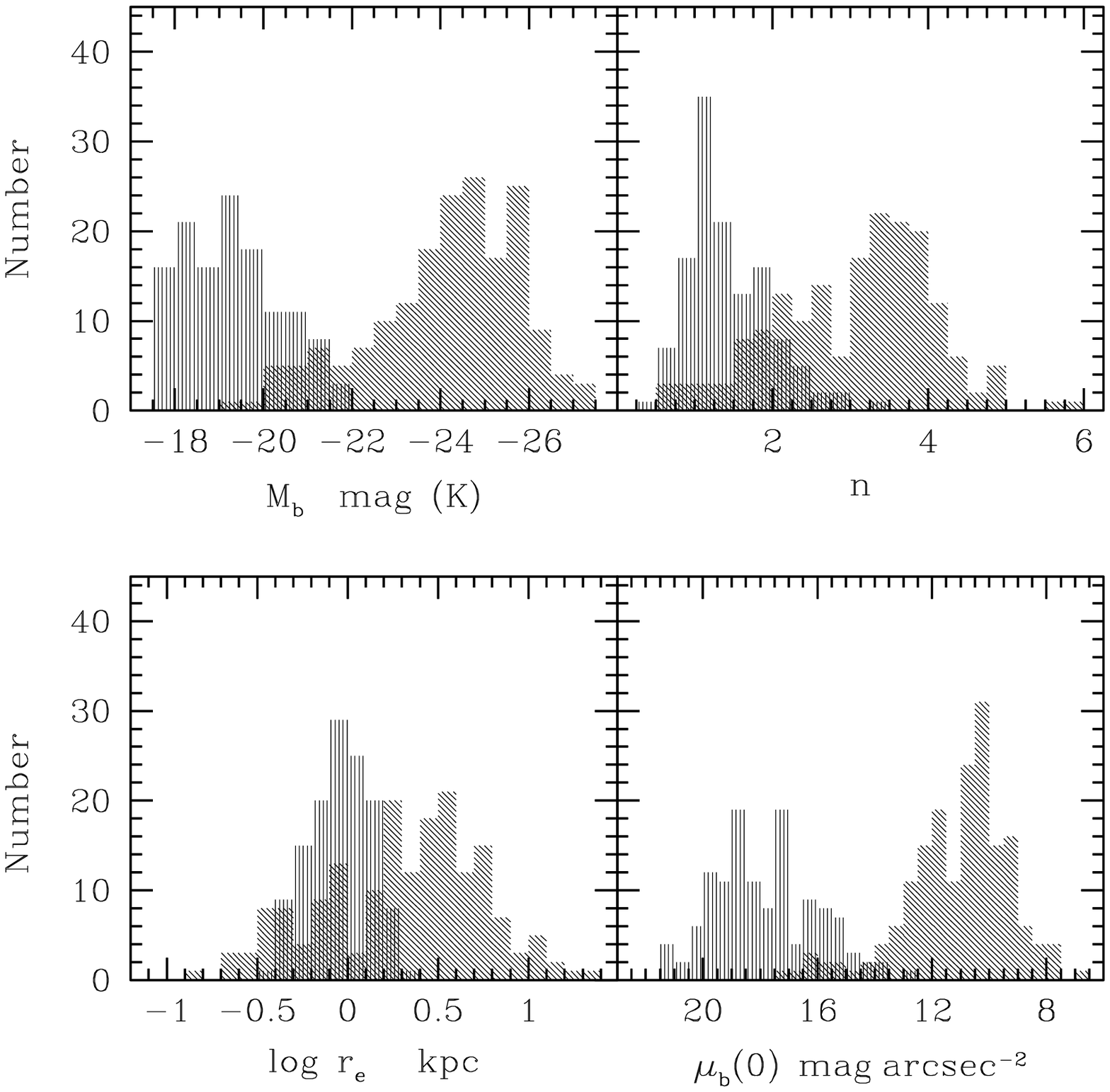}{Combined histogram for the structural properties 
of bulges in our sample. There seems to exist two distinct groups in 
their structural properties, irrespective of the Hubble type they 
belongs to.}

\subsection{Correlations}
        The three basic parameters involved in the S\'ersic law given in  
\equn{bint}, $\ibzero$, $\re$ and $n$, specify the surface brightness 
profile completely. $\ibzero$ determines how bright the galaxy appears, 
$\re$ decides 
the extent of the galaxy and $n$ controls the `rate' of the fall of
intensity within the galaxy;
the larger the value of $n$, the quicker is the fall in the intensity
within the half light radius. 
So for large values of $n$, it is necessary to have a bright core 
and a reasonably large $\re$ for the galaxy to be `visible' in the band.
Hence positive correlations (with some scatter) of $n$ with 
both $\ibzero$ and  $\re$ are expected, assuming galaxies obey the 
S\'ersic law over the entire galaxy (\eg D'Onofrio, Capaccioli, 
\& Caon 1994; Khosroshahi \etal 2000b; Trujillo \etal 2002).

 In Fig.~\ref{mu0n} we show a plot between $\mubzero$ and $n$ for the 
entire sample. In the case of the cluster ellipticals, there is a good linear 
correlation (correlation coefficient = -0.90, significance greater than 
$99.99\%$, rms 
scatter around the best fit line measured along $\log n$ axis = 0.26), The 
dwarf ellipticals too 
show a  good correlation, but there is pronounced departure from linearity 
at the lower values of $n$.  The dwarf galaxies seem to lie on the 
extrapolation of the trend in ellipticals to lower $n$ values, 
 but this is to be treated with some 
caution, since we have used an average color to transform the observed 
dwarf galaxy magnitudes in the $B$ band to the $K$ band. As for the bulges 
of spirals, early type bulges again show a good correlation ($-0.88$, 
$>99.99\%$) and merge with 
ellipticals (see also Khosroshahi \etal 2000b), while late type bulges show 
a large scatter and departure from 
linearity. The bulges of  brighter lenticulars mix  with the ellipticals, 
while their fainter counterparts show large scatter and curvature.

\plotn{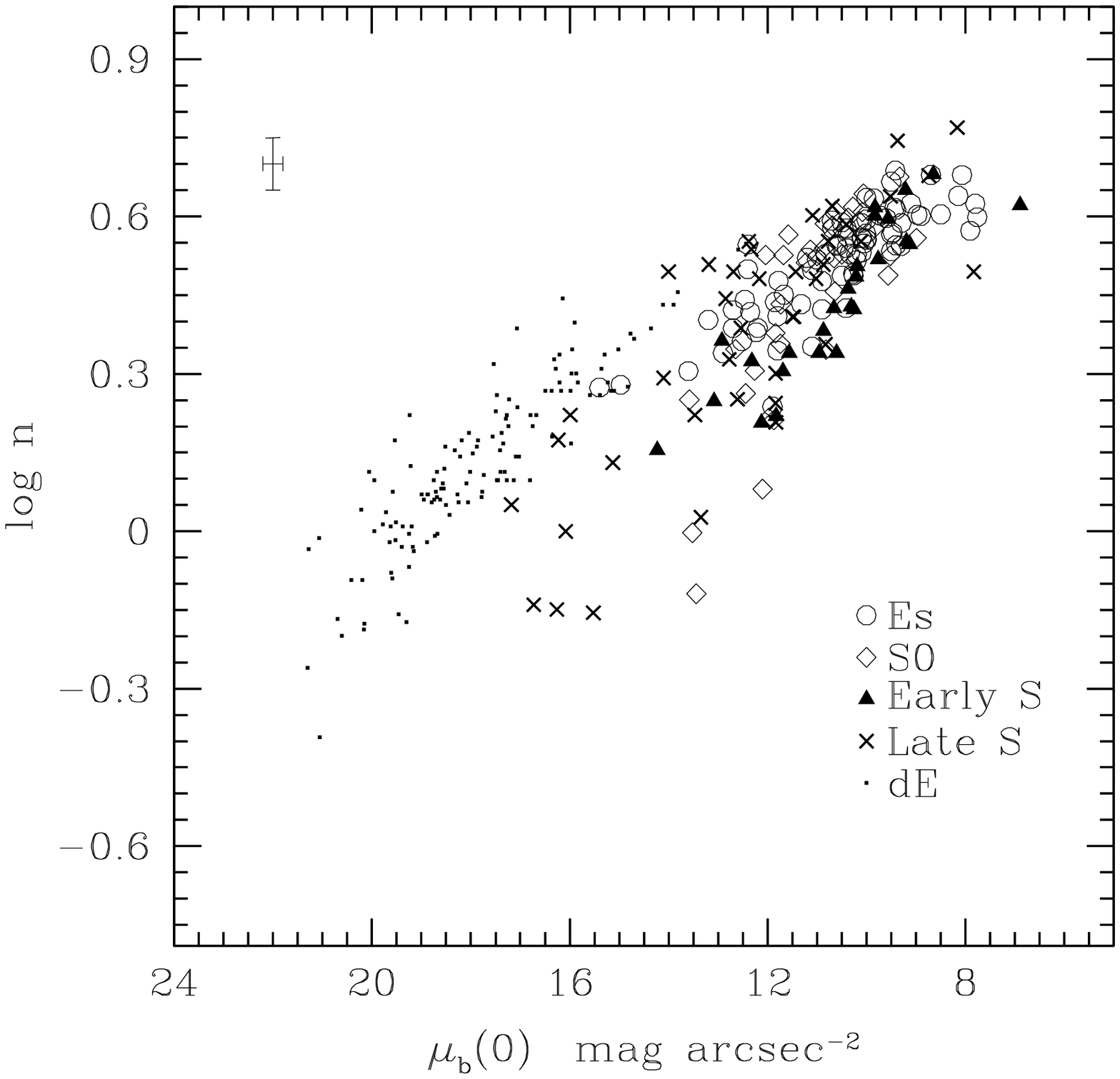}{Logarithm of the S\'ersic index $n$ as a function of 
unconvolved bulge central surface brightness.}

A plot of the  S\'ersic index against the effective radius (Fig.~\ref{nre}) 
shows the presence of two broad distributions, but now without a good 
correlation within each group. It is seen from the figure that 
Es and brighter bulges of 
other Hubble types form a group, while the fainter (and smaller) bulges 
of S0s and spirals form another group similar to that of dEs.
\plotn{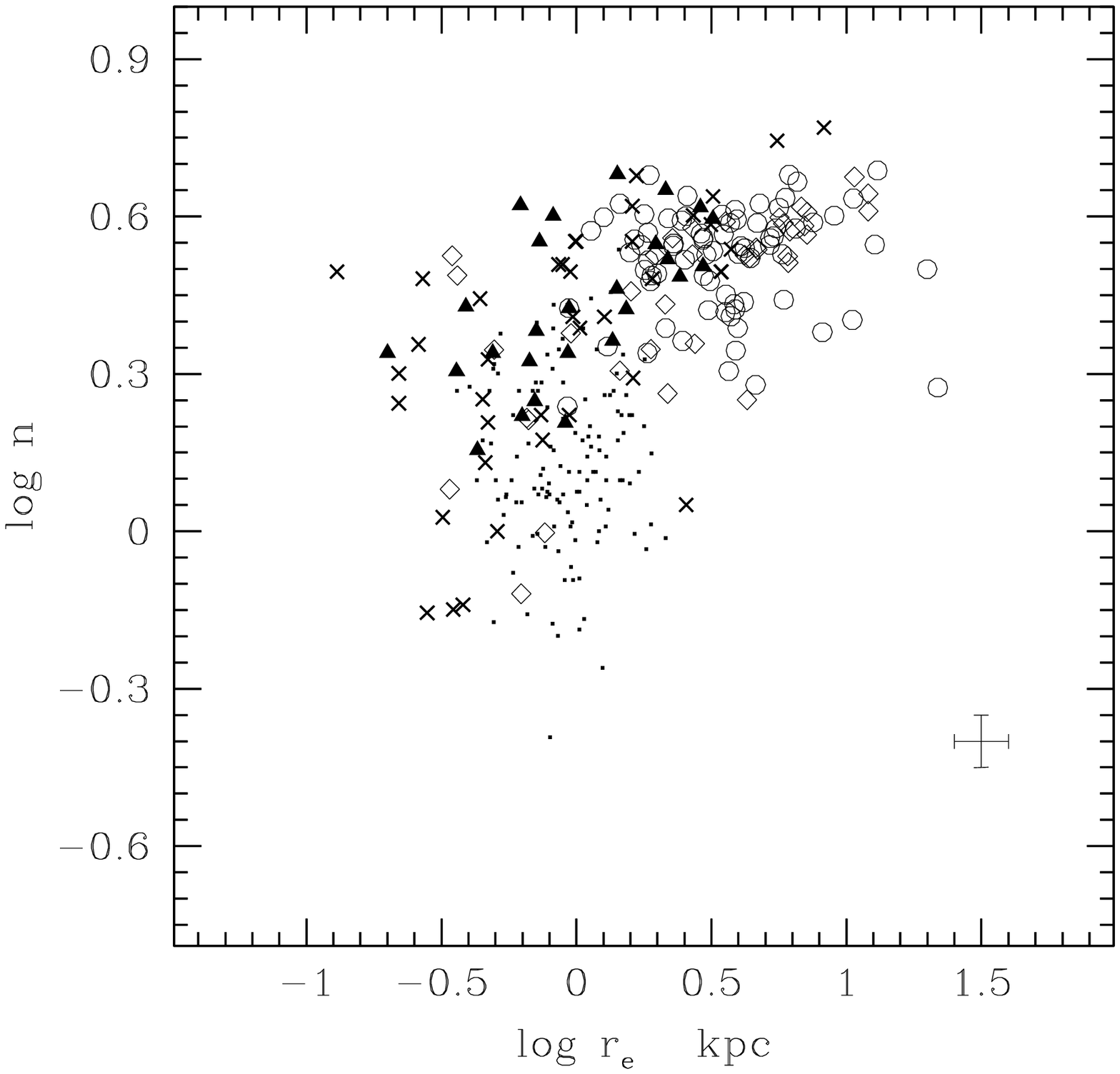}{S\'ersic index Vs effective radius. The symbols are as used 
in Fig.~\ref{mu0n}. Typical error bars are also shown at the lower
right corner.}
\subsection {Kormendy Relation}
Kormendy (1977) noticed that for a sample of large elliptical
galaxies, the central surface brightness $\mubzero$ was correlated
with $\log \re$.  The two parameters here were determined by fitting
de Vaucouleurs' law to the SB distribution.  Operationally, it has
been found in such investigations that it is convenient to use the
mean surface brightness within $\re$, $\msbe$ rather than $\mubzero$
in plotting correlations like the Kormendy relation and the
fundamental plane, as the former can be more robustly determined from
observation data, and the two differ from each other by just a
constant when de Vaucouleurs' law is used : $\msbe = \mubzero + 6.935$
for $n=4$. However, care should be taken when results from the 
  $\dev$ fitting are compared with those from $\ser$ fitting, as the
  photometric structural parameters are affected systematically. 
  In the case of Abell cluster ellipticals,
  where we have both  S\'ersic and de Vaucouleurs law fitted parameters,
  the propagation of the systmatic error is shown in Fig.~\ref{n_rat}. It
  can be seen that the more different $n$ is from $4$ for the bulges, the
  larger are the  systematic errors involved in the de Vaucouleurs'
  fitting. However, there was no noticeable difference for the
  Kormendy relation ($\msbe  - \re$) between the two cases, as the
  systematic errors were 
  propagating  similarly along both axes (Ravikumar 2004, see also
  Kelson \etal 2000).
\plotn{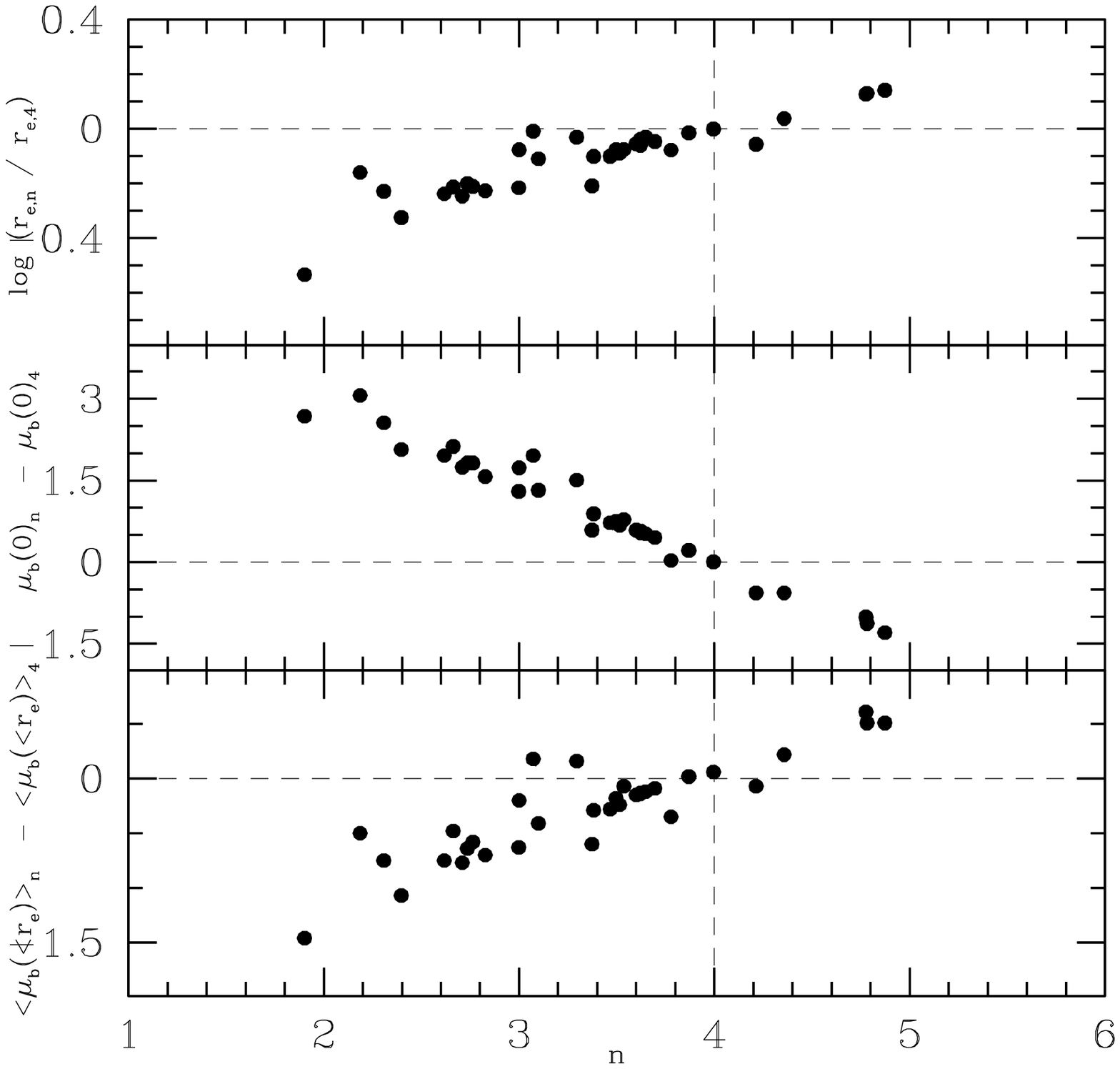}{The systematic error involved, when the de Vaucouleurs
  law is used to extract bulges with a  S\'ersic profile. Fig.~\ref{n_rat}.}
\plotn{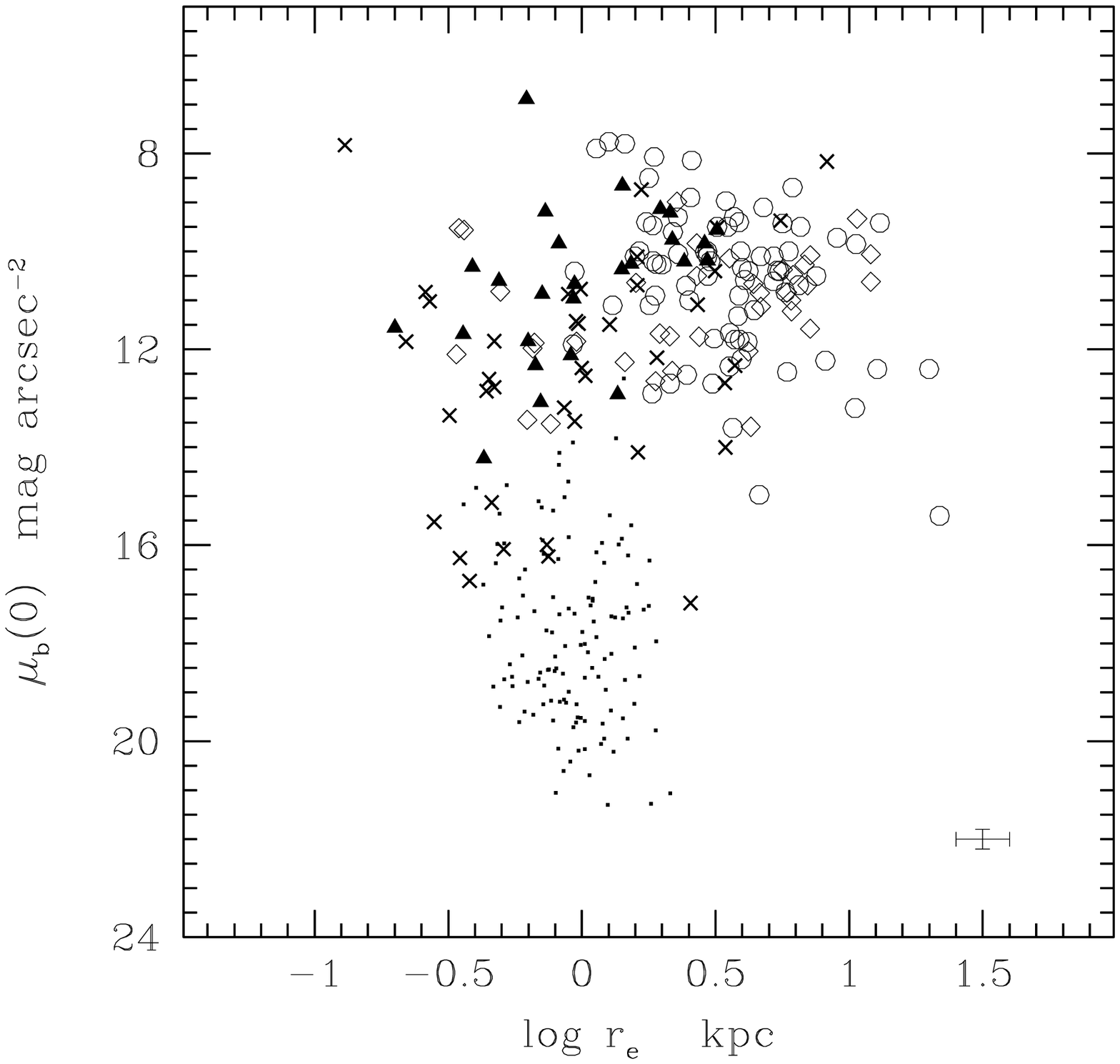}{Variation of $\mubzero$ as a function of effective radius. 
The symbols are as used in Fig.~\ref{mu0n}. Typical error bars are
also shown.}

   In Fig.~\ref{mu0re} we have shown a plot of $\mubzero$ against $\log \re$. 
The correlation is not good, but here again we see the various kinds of 
bulges are divided into two distinct groups, as in Fig.~\ref{nre}. The lack 
of correlation seems to go against the Kormendy relation, but the 
correlation is clearly visible if we plot $\msbe$ against $\log \re$, 
as shown in Fig.~\ref{mkor}.  When the S\'ersic law is used the relation 
between the two surface brightness terms is  a function of $n$, 
and the lack of a tight $\mubzero$-$\log \re$ correlation, is usually 
attributed to the interdependence of $n$ with these parameters separately 
(see Khosroshahi \etal 2000b and Graham \& Guzm\'an 2003).

\begin{table*}
\begin{center}
\caption{Kormendy relation,$\msbe = a \log \re + b$.  The linear 
correlation coefficient $r$ with Significance are also given.} 
\label{tab:kor}
\begin{tabular}{lcccc}\hline\hline
Sample  & $a$   &  $b$ & $r$ & Significance \\
\hline
Es         &1.95  $\pm$ 0.31   &15.31   $\pm$ 0.17   & 0.71  & $>99.9$  \\
S0s        &1.46  $\pm$ 0.49   &15.34   $\pm$ 0.19   & 0.90  & $>99.9$  \\
Early S    &2.60  $\pm$ 0.32   &16.86   $\pm$ 0.15   & 0.48  & $~~~98.7$  \\
Late  S    &2.53  $\pm$ 0.22   &15.14   $\pm$ 0.17   & 0.74  & $>99.9$  \\
dE         &0.46  $\pm$ 0.38   &13.64   $\pm$ 8.14   & 0.31  & $>99.9$  \\
\hline
\hline
\end{tabular}
\end{center} 
\end{table*}

        It appears at first sight from Fig.~\ref{mkor} that there is large scatter in the Kormendy relation for our bulges, but it is clearly seen that the points are segregated into well correlated subsets. We have given in \tablem{kor} the correlation coefficients and best fit values for the Kormendy relation for the different kinds of bulges. It is seen that the slope for the dEs is significantly different from that for other sets. The errors in the best fit values have been obtained using the bootstrapping and unless otherwise mentioned, we use this re-sampling technique for estimating errors in the rest of the paper. 

In Fig.~\ref{mkor} we have shown the line corresponding to  $M_{\rm K} = -22$ which separates the whole sample into two groups which separately obey a tighter correlation than the combined sample. The slope of the relation for bulges in the brighter group is higher than the slope of the fainter group. Khosroshahi \etal (2004) also reported such a segregation of bulges of early type galaxies in the magnitude space with different slopes for the Kormendy relation. We note  that while the S0s and bulges of early type spirals obey the Kormendy relation for the Es in our sample, the bulges of late type spirals form a bridge between the two groups. 

        Kormendy (1985) also noticed such a `discontinuity' between Es and dEs in the parameter correlations. Interestingly, Fig.~\ref{mkor} shows that the fainter bulges of late type spirals are more similar to the dEs in their slope while their brighter counter parts show a weak similarity with the bulges of E/S0s. Recent studies suggest that bulges of late type spirals are more elongated than those of early type spirals (Fathi  \& Peletier 2003). Combining this with the fact that dEs have greater rotational support than the Es, it is tempting to see the Kormendy relation as an evolutionary diagram but there is too much mixing between the different types, for any clear trend to be discerned (see also Graham \& Guzma\'n 2003).

\plotn{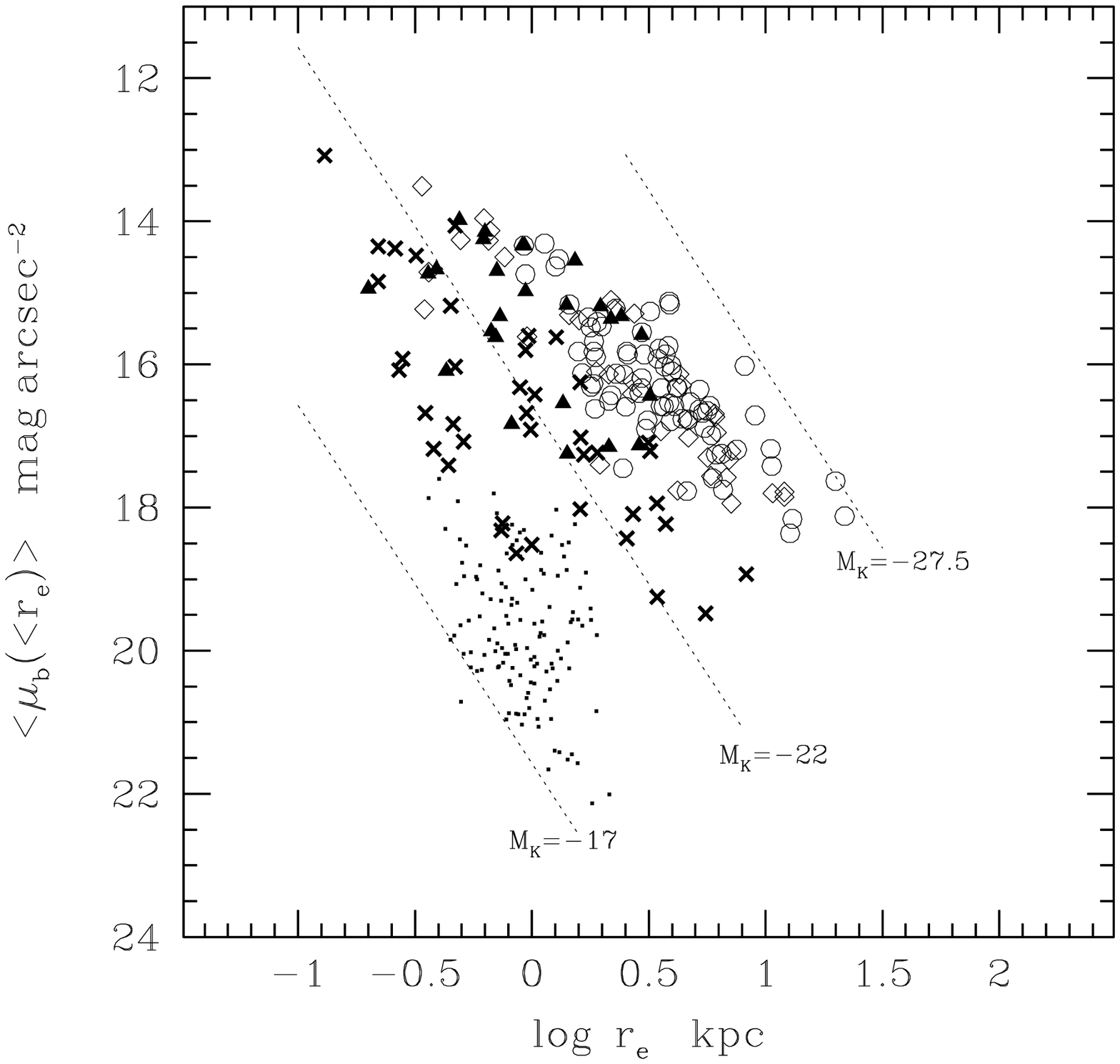}{Kormendy relation for bulges in our sample. The (dashed) 
lines of constant absolute magnitude are also shown.}
\subsection {Correlations with total luminosity}
In Fig.~\ref{totmc} we show the correlations of structural parameters
of the bulges in our sample with the total luminosity of the host
galaxies. The two surface brightness terms ($\mubzero$ and $\msbe$)
seem to correlate well with the total magnitude, even though with
large scatter, while for galaxies other than the dEs the half-light
radius also shows significant correlation. 

\plotn{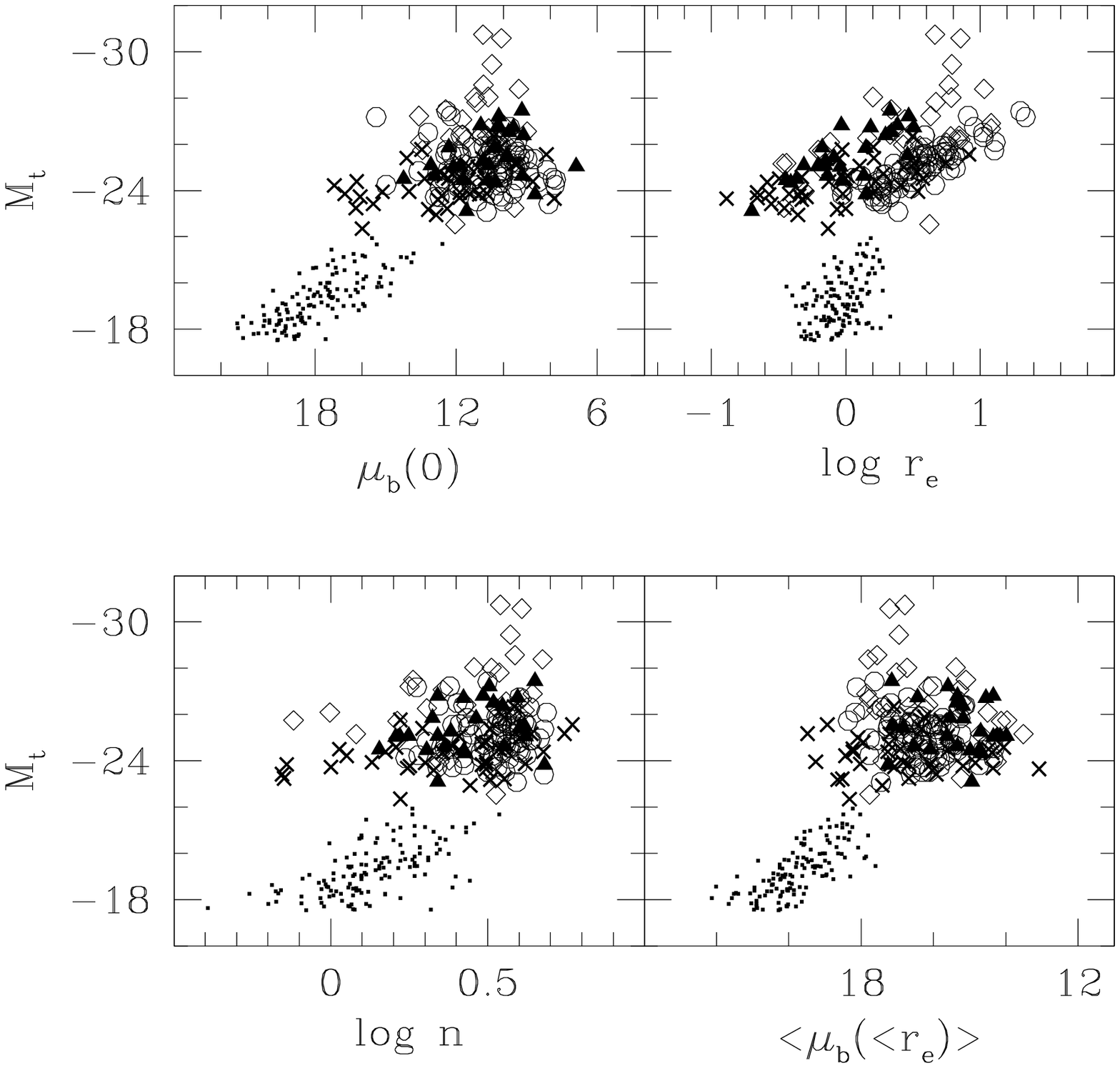}{Correlations of bulge structural parameters $\mubzero$
  (in $\magper$), $\re$ (in kpc), $n$, and $\msbe$ (in $\magper$)  with total
  $K$ magnitude for different galaxies in our sample. For the cluster
  ellipticals and dEs in our sample, the total luminosity is the same
  as that of the bulge, as no significant disks were detected in
  these systems.}
\subsection {The Photometric Plane}
   Khosroshahi \etal (2000a) noticed that elliptical galaxies, and the bulges 
of early type spiral galaxies satisfied a single planar relation of the 
form $\log n = a \log \re + b \mubzero +c $. The data points had small 
scatter around the best fit relation, which Khosroshahi \etal (2000a,b) 
called the \emph{photometric plane}. They argued that the tight relation 
suggested that there is
a single formation scenario for the ellipticals and early type bulges. 
A less direct form of this relation can be seen in Binggeli \& Jerjen (1998) 
for a sample of dwarf ellipticals and dwarf S0s in the Virgo cluster. 
For their sample of galaxies containing mostly late type spirals 
M\"ollenhoff \& Heidt (2001) noticed a similar relation for 
the S\'ersic parameters obtained  by combining  observations 
in near infrared ($J$, $H$, and $K$)  bands. Khosroshahi \etal (2000a) 
suggested that the PP can be used in distance determinations, albeit 
with  uncertainties as large as $\sim50\%$.  
Later, Graham (2002) obtained a slightly different form of the PP, 
from $B$ band observations of early type galaxies 
in the  Virgo and Fornax clusters, and used it to estimate the Virgo-Fornax 
distance modulus. 
A key difference between
the photometric plane and the fundamental plane 
(where the distance estimation error can be 
 $\sim 10\%$ in rich clusters) is that in the former, all the parameters 
are photometrically determined, which can be an advantage for large 
samples of distant galaxies with no spectroscopic data.  

\begin{table*}
\begin{center}
\caption{The photometric plane coefficients. The number of galaxies 
in the set ($N$) and the angle between the normal of the PPs of each set 
and that of the Es are also given.}
\label{tab:pp}
\begin{tabular}{lcccrcc}\hline\hline
Sample  &  $a$   &  $b$ & $c$  & N & $\unit{rms_{n}}$ & Angle \\
\hline
Es         & 0.153$\pm$0.022  &-0.066$\pm$0.003  &1.13$\pm$0.03   & 76 & 0.037 &  -  \\
S0s        & 0.206$\pm$0.030  &-0.106$\pm$0.014  &1.54$\pm$0.15   & 37 & 0.067 &   3.695$\pm$1.620  \\
Early S    & 0.130$\pm$0.040  &-0.073$\pm$0.010  &1.21$\pm$0.11   & 26 & 0.058 &   1.354$\pm$1.777  \\
Late S     & 0.232$\pm$0.029  &-0.074$\pm$0.006  &1.29$\pm$0.07   & 40 & 0.092 &   4.391$\pm$1.912 \\
Spirals    & 0.213$\pm$0.023  &-0.067$\pm$0.005  &1.18$\pm$0.06   & 66 & 0.088 &   3.346$\pm$1.617 \\
dEs        & 0.158$\pm$0.035  &-0.082$\pm$0.004  &1.60$\pm$0.06   & 128& 0.071 &   0.965$\pm$1.202 \\
\hline
\hline

\end{tabular}
\end{center} 
\begin{list}{}{}
\item Note:~~$a$ and $b$ are the coefficients of $\log \re$ and 
$\mubzero$ respectively, $c$ is the constant and $\unit{rms_{n}}$ is 
the rms scatter in the PP measured along the $\log n$ axis.
\end{list}
\end{table*}

    In this section we examine the PP relations for the various kinds of 
bulges  of galaxies 
in our sample.  The best fit relations are given in \tablem{pp}, where in 
addition to the coefficients $a$, $b$, and $c$, we provide the rms scatter 
along the $\log n$ axis and the  angles made by the normal to the PP for each 
group with that for the Es. An edge-on view of the PP for ellipticals is 
shown in Fig.~\ref{pp_e}; these galaxies show a remarkable homogeneity and have 
significantly smaller rms scatter, about the best fit plane, than the other 
subsets. The PP coefficients for the bulges of  early type spirals and 
ellipticals are equal, within $1\sigma$ errors, as shown in \tablem{pp}.  
The PPs for bulges of S0s and late type spirals  are, however,  
significantly different from the PP for ellipticals, as can be seen 
from the coefficients and the angles between the normals. The scatter 
about the PP in these cases is also larger than that for the Es.  
The sample of dwarf galaxies  is some what unique in their properties 
in the sense that they show a PP very similar to that for Es, but with 
much larger scatter, implying that a  high degree of inhomogeneity is 
present in the individual galaxies.  

\plotn{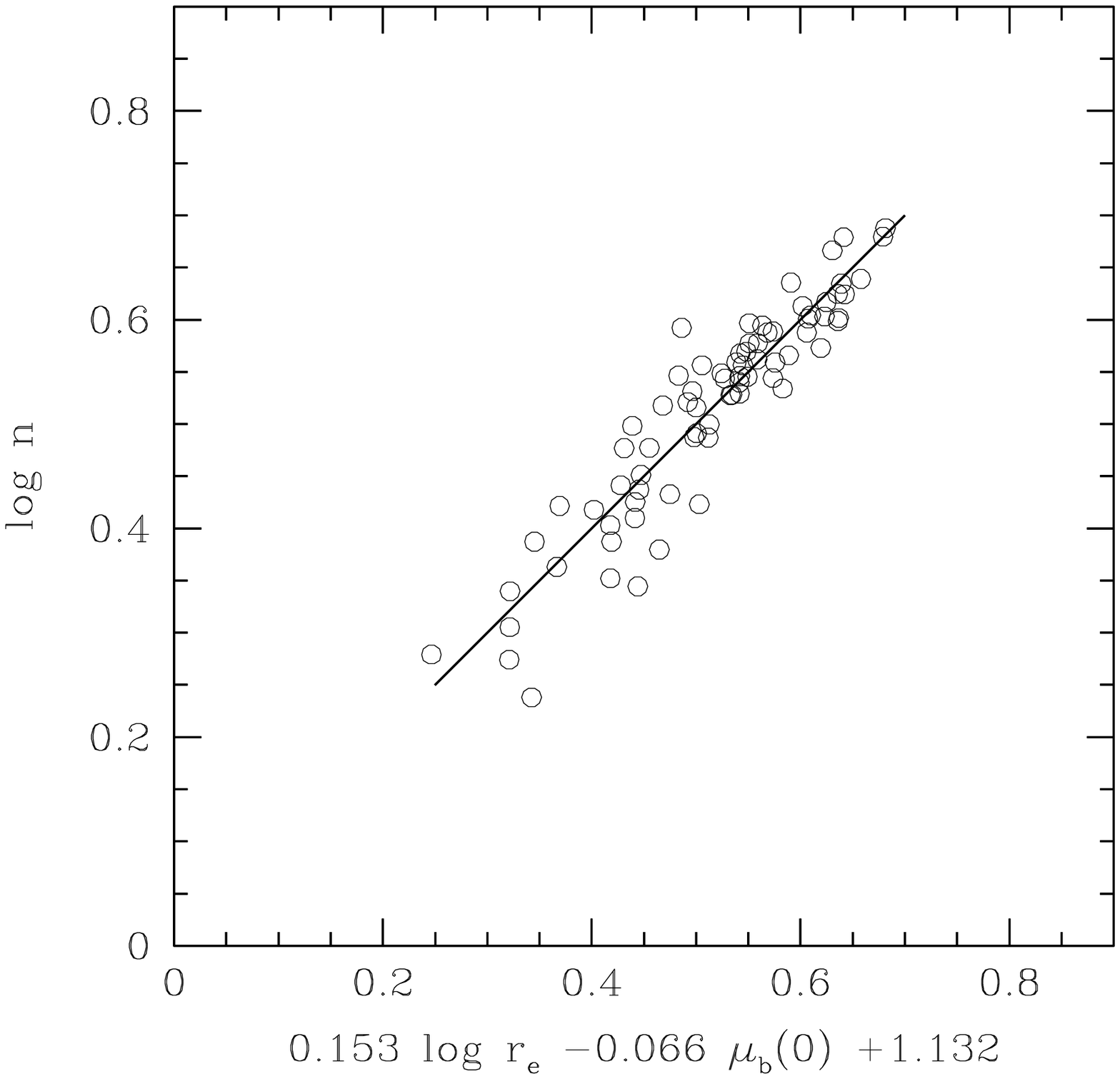}{The Photometric Plane for Es; an edge-on view. 
The different sets of cluster ellipticals have a tight distribution
about the PP.}

        The PPs for types other than the Es are shown in Fig.~\ref{pp_os}.  
Three points are noteworthy here. First,  not all bulges follow 
a common PP. Second, the ellipticals and bulges of early type spirals form a very homogeneous
and tight PP while all other samples show a curvature towards the lower 
end of $n$, in confirmation with recent results (Khosroshahi \etal 2004). 
Finally, of the two groups of bulges seen in  Fig.~\ref{cuhist}, the bigger 
and brighter bulges (with larger $n$ values) form a PP with much less scatter 
than the fainter ones. 
\plotn{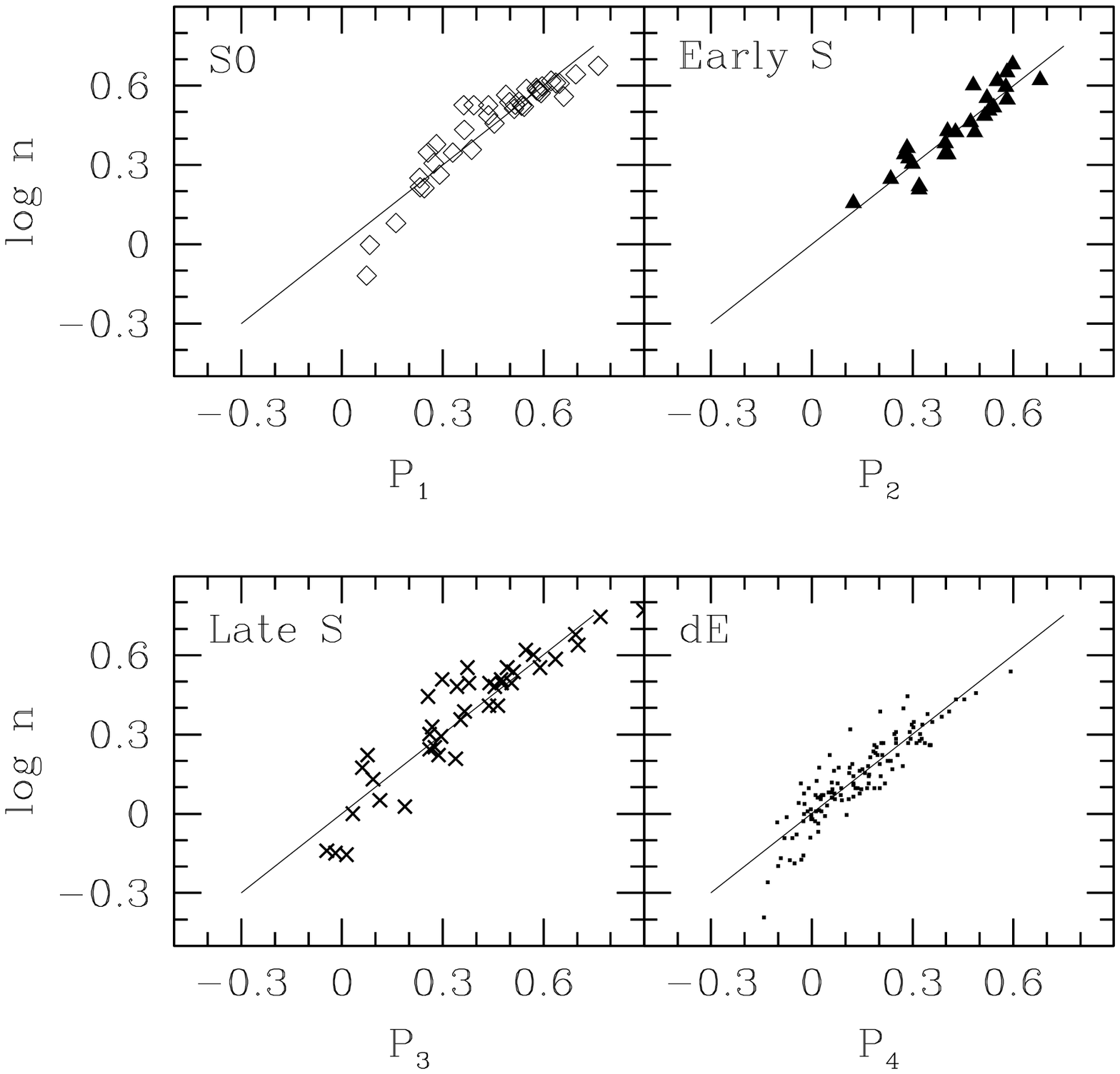}{Edge on view of PPs for S0s (top left window), bulges of early 
(top right) and late (bottom left) type  spirals, and dwarf galaxies 
(bottom right). $P_1 = 0.21 \log \re -0.11 + 1.54$, $P_2 = 0.13 \log \re -0.07 + 1.21$, $P_3 = 0.23 \log \re -0.07 + 1.29$, and  $P_4 = 0.16 \log \re -0.08 + 1.60$}
\plotn{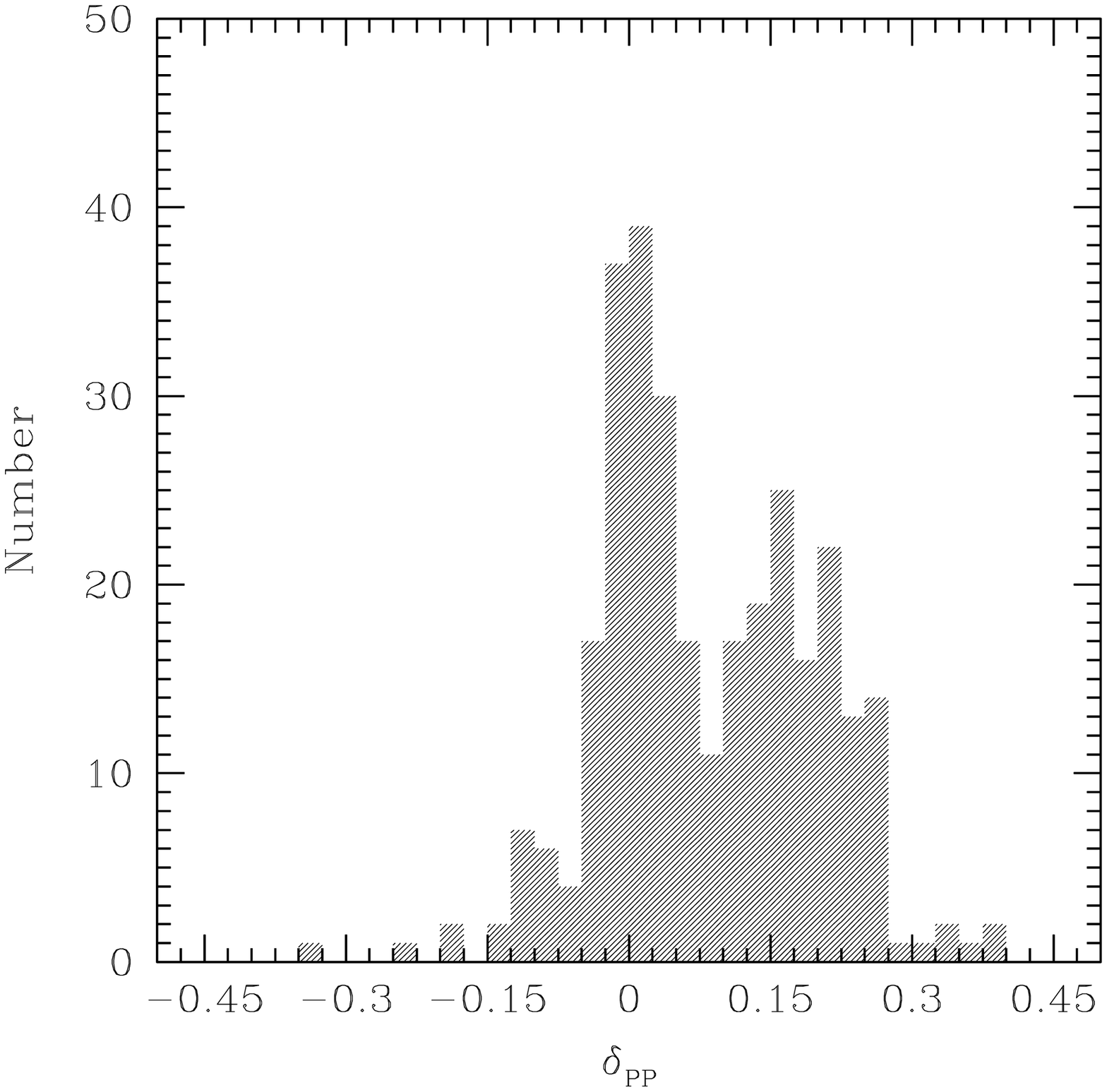}{Histogram of absolute deviations from the PP defined by 
ellipticals,
  $\delta_{\rm PP} = \log n - (0.153 \log \re -0.066 \mubzero + 1.132)$,
for the bulges in the analysis. The presence of two distinct groups is 
clearly visible.}
        The dichotomy is clear in Fig.~\ref{hdpp} where we show the 
histogram of deviations from the PP defined by Es for the combined sample. 
We have plotted in Fig.~\ref{dppcor} 
the absolute deviation for each galaxy in our sample from the
photometric plane for the Es. There is no noticeable trend in the 
case of $\log n $ and $\log \re$,  possibly  because of the significant 
overlap between the brighter and fainter groups in the distribution
of these  parameters.   However, the $|\delta_{\rm PP}|$ show significant 
correlation with the central intensity and luminosity  of the bulges, 
with the average deviation increasing systematically as these values 
become fainter.

\plotn{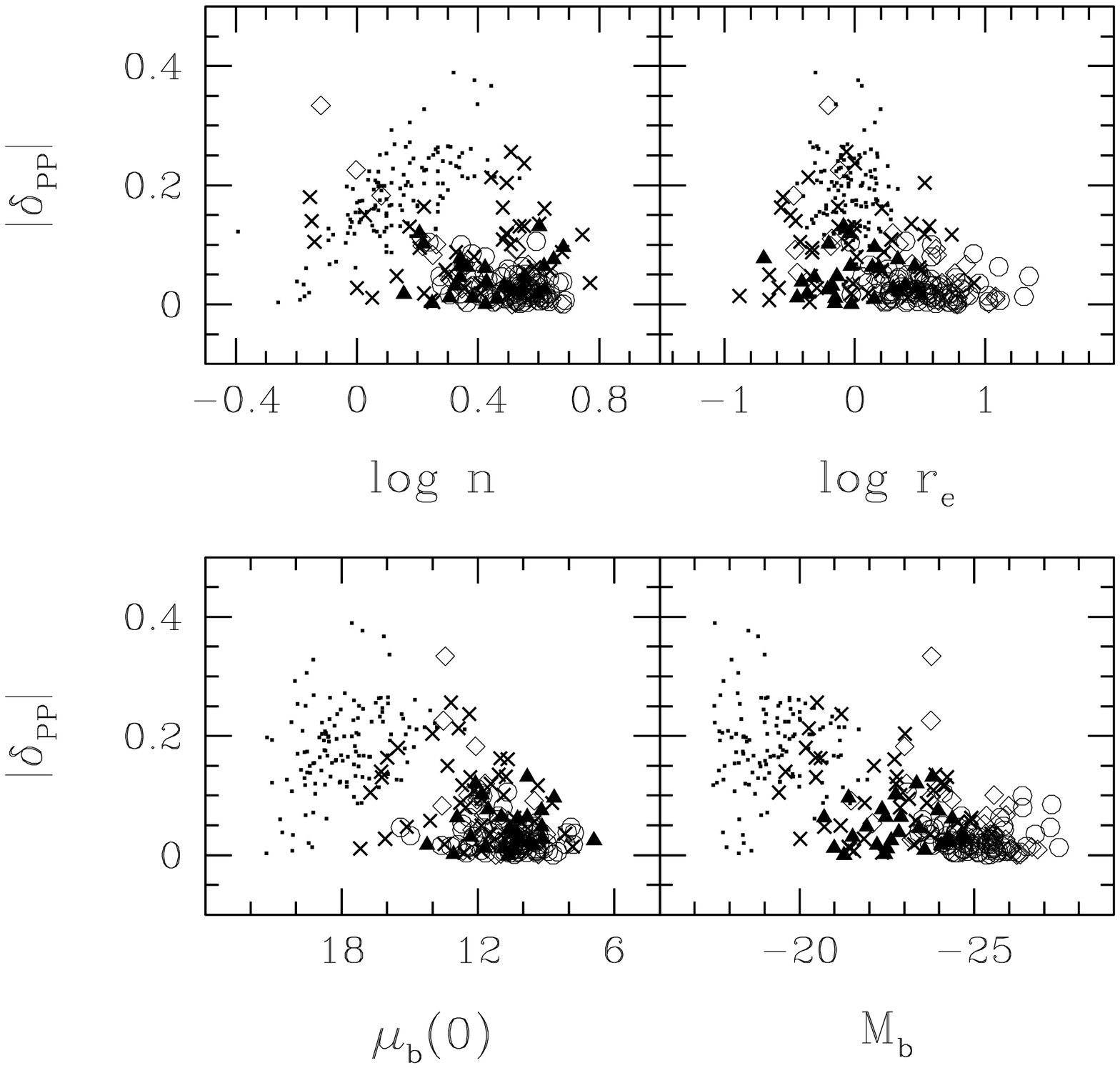}{The variation of $|\delta_{\rm PP}|$ as a function of 
the bulge parameters $n$, $\re$ (in kpc), $\mubzero$ (in $\magper$), 
  and  $M_{\rm b}$ (in $K$ mag). The brighter the central intensity or luminosity
 of the bulge, the higher the chances that it falls on the PP.} 

To examine  the distribution of bulges in the 2-dimensional space 
defined by the photometric plane of the Es,  we show a face on view 
of the PP in Fig.~\ref{fa_e}. The axes ${\bf K_1} = 0.151 \log n + 0.989 \log \re$ 
and ${\bf K_2} = -0.064 \log n + 0.010 \log \re-0.998 \mubzero $ form 
orthogonal triad with the normal to the PP for Es. Clearly, the whole available 
space in the PP is not occupied and, there is suggested distribution of 
different Hubble types according to their luminosity. There appears to be a 
(continuous) sequence from the dEs to Es, through the other Hubble types. 
Elliptical galaxies are considered to be the most relaxed systems, which is 
consistent with their tight distribution around the FP and PP, and we have
seen that the dEs have greater scatter around the PP than other types of
galaxies.  So the sequence may be tracing a trajectory in the 
${\bf K_1/K_2}$ plane, which goes from the least to the most relaxed systems. 
In the next section we try to relate this sequence to the specific 
entropy of bulges (see Merritt 1999).  
\plotn{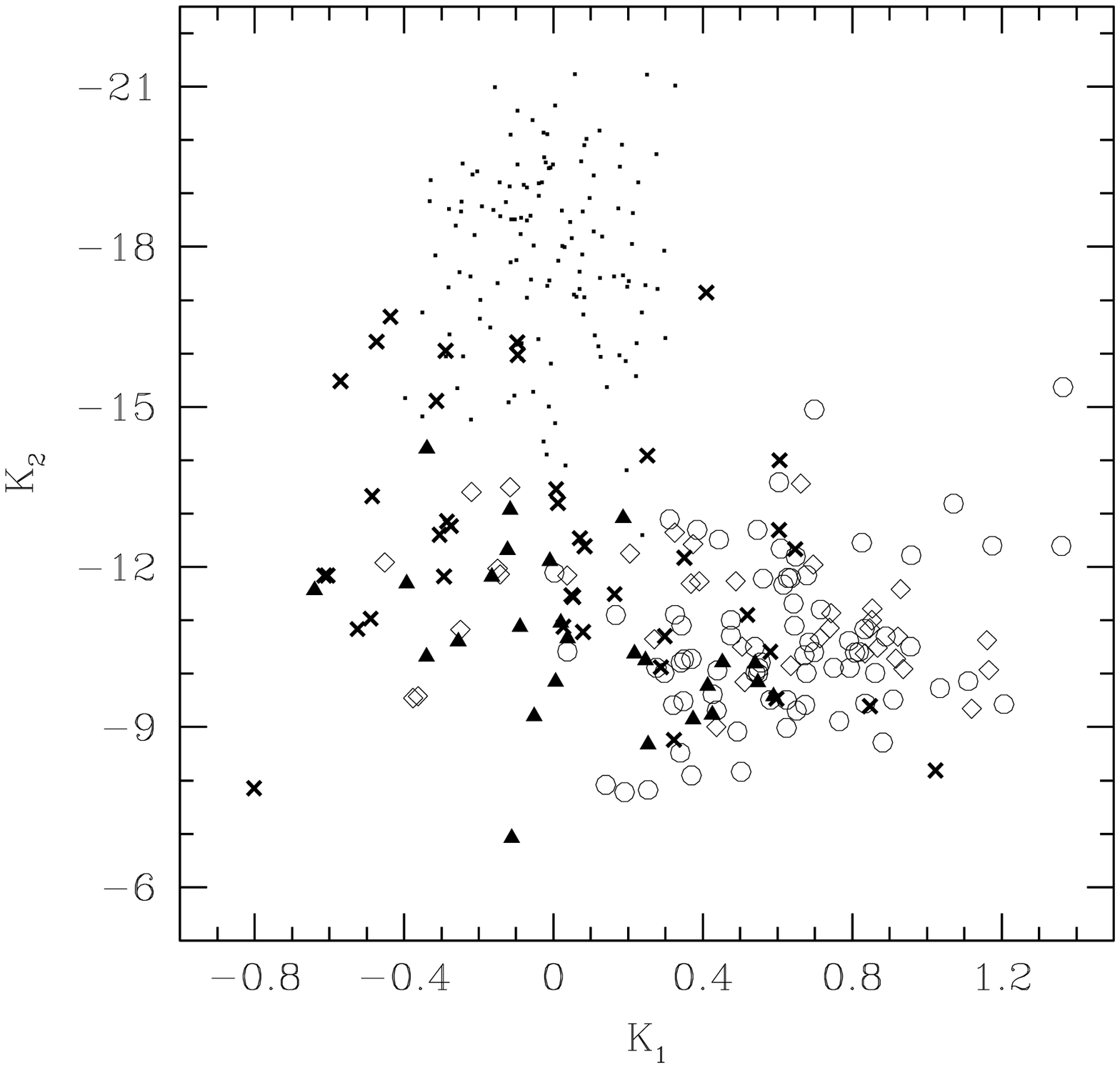}{Distribution of bulges on the PP for Es. The PP for Es is 
shown face on. The axes are $K_1 = 0.151 \log n + 0.989 \log \re$ and
 $K_2 = -0.064 \log n + 0.010 \log \re-0.998 \mubzero$.}

\section{The specific entropy of bulges}
     Galaxies are in a state of constant evolution. However, studies 
on galaxies of a particular Hubble type show that they indeed display 
remarkable similarities in their structural properties, manifested by 
various correlations, photometric and spectroscopic, like the FP and 
Faber-Jackson relations for the early type galaxies and Tully-Fisher (1977)
relation for spirals. The existence of such regularities has  
interpreted to mean that different classes of galaxies are in a quasi-
equilibrium state of their own. Assuming that there exists a local 
maximum entropy associated with each quasi-equilibrium state, it is possible 
to estimate the entropy of a self gravitating system, from thermo-dynamical 
considerations.  Considering the stars in an elliptical galaxy to behave 
like an isolated, self-gravitating gas, Lima Neto \etal (1999) suggested 
that the Es might be treated as systems of constant specific entropy.
Later, M\'arquez \etal (2000) 
showed that the specific entropy of galaxies, estimated from the 
microscopic Boltzmann-Gibbs definition, increases with
merging history. In the latter approach,  stars in the galaxies are 
not required to obey the ideal gas law. Taking their approach 
further, M\'arquez \etal (2001) have suggested a theoretical
explanation of photometric properties of the Es, including the PP,
by constructing an \emph {energy-entropy line}. 

        The different evolution scenarios for galaxies can be 
treated as processes associated with continuous increase in entropy 
through quasi-stationary states. In  the case of hierarchical merging
(Kauffmann \& White 1993) or where the violent 
relaxation mechanism (Lynden-Bell 1967) operates, the phase-space mixing is 
expected to be fast and hence would result in a higher quantum of 
change in entropy. On the other hand, in the case of secular evolution 
(Kormendy 1979) the change in entropy, one may expect,  is 
rather slow, as the involved time scales are large.

Following the formulation of M\'arquez \etal (2000; equation 2), the 
specific entropy of a spheroidal system can be written as, 
\myequn{smarq}{S = 0.5 \ln \ibzero + 
2.5 \ln \frac{\re}{\sib{1.999n - 0.327}}+F(n),}
where $F(n) = 3.9 n^{-1.3} -1.3 n -0.2 \ln n$.

It should be noted that we have modified the original M\'arquez 
\etal (2000) equation by representing the scale radius in \unit{kpc} 
(instead of angular units) and neglecting the constant,  
in order to have a comparison of specific entropy estimates of galaxies 
in our sample. 
In \tablem{entp} we show the mean values of specific entropy estimated for the 
bulges of different Hubble types in our sample. The Es and dEs  show a 
tighter distribution in specific entropy, evident from their dispersion 
values,  suggesting that these structures have  higher stability 
relative to the others. It is clear that the 
specific entropy decreases systematically as we go from Es to dEs through S0s 
and spirals. This supports a merging scenario hierarchical or violent 
relaxation of dEs which 
then successively climb up in the entropy ladder to form Ellipticals which 
top the table in specific entropy.

\begin{table*}
\begin{center}
\caption{Mean Specific Entropy for Bulges in Our Sample.}
\label{tab:entp}
\begin{tabular}{lc}\hline\hline
Sample  & $S$    \\
\hline
Es         & 18.39  $\pm$ 4.44   \\
S0s        & 16.04  $\pm$ 7.35   \\
Early S    & 12.54  $\pm$ 6.22   \\
Late  S    & 12.48  $\pm$ 8.32   \\
Spirals    & 12.50  $\pm$ 7.51   \\
dE         & ~8.44  $\pm$ 1.98   \\
\hline
\hline
\end{tabular}
\end{center} 
\end{table*}

\plotn{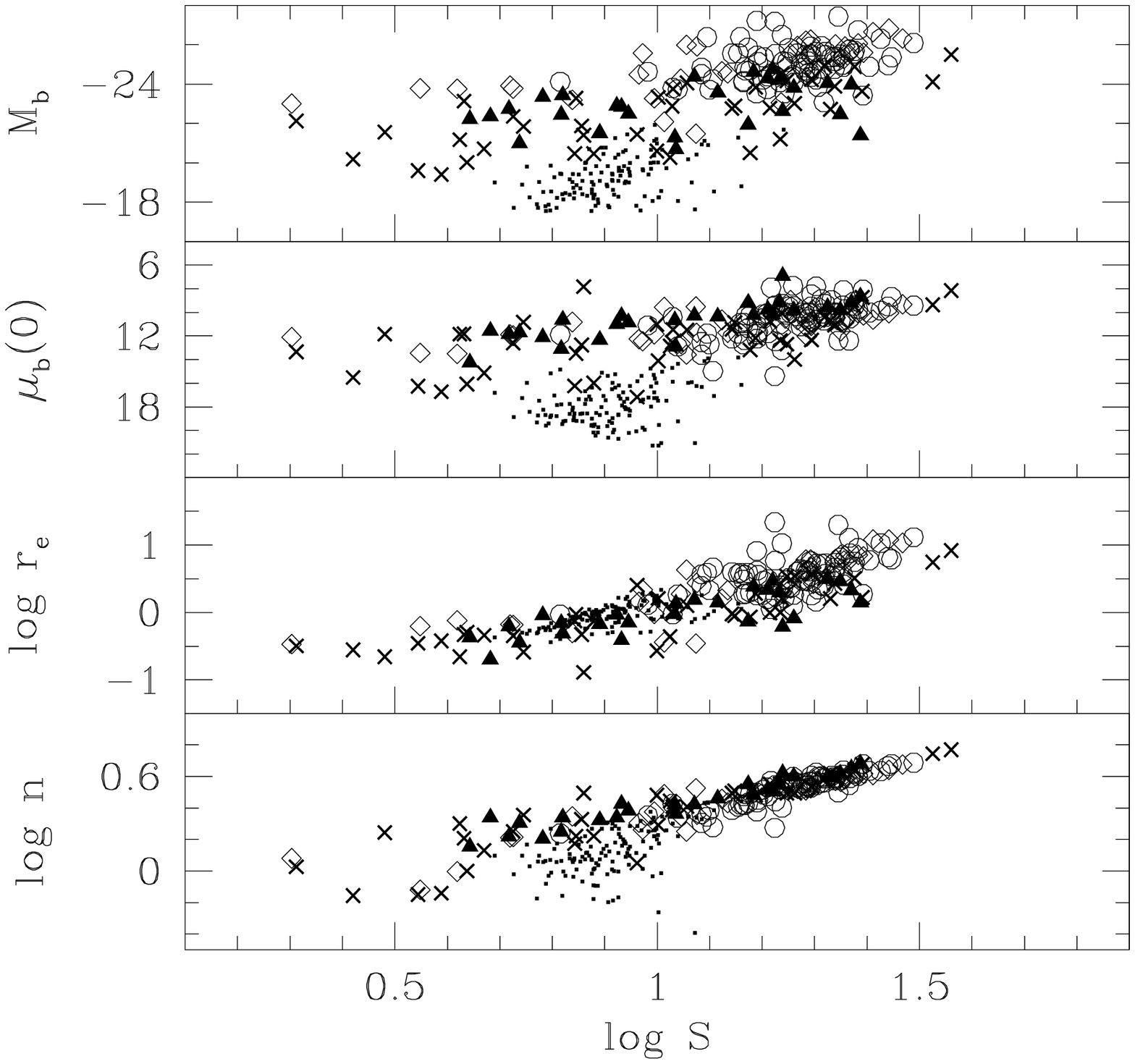}{Variation of $n$, $\re$ (in kpc), $\mubzero$ (in
  $\magper$), 
  and  $M_{\rm b}$ (in $K$ mag)  as a function of specific 
entropy.}
        In Fig.~\ref{entp} we show the variation of specific entropy
with the bulge parameters. Clearly there is a systematic increase in
the entropy as the bulge becomes brighter and bigger.  The
distribution of specific entropy of bulges on the photometric plane
also shows a systematic pattern as shown in the Fig.~\ref{face}. As
the luminosity (and hence the mass) of the galaxy is proportional to
the central intensity and effective radius squared, the observed trend
of specific entropy support a scenario where the less massive systems
merge to form more massive ones. The correlation between the
 S\'ersic index and the specific entropy seems to be the tightest,
suggesting that the index may reflect the level of interactions
the galaxy has passed through.

\plotn{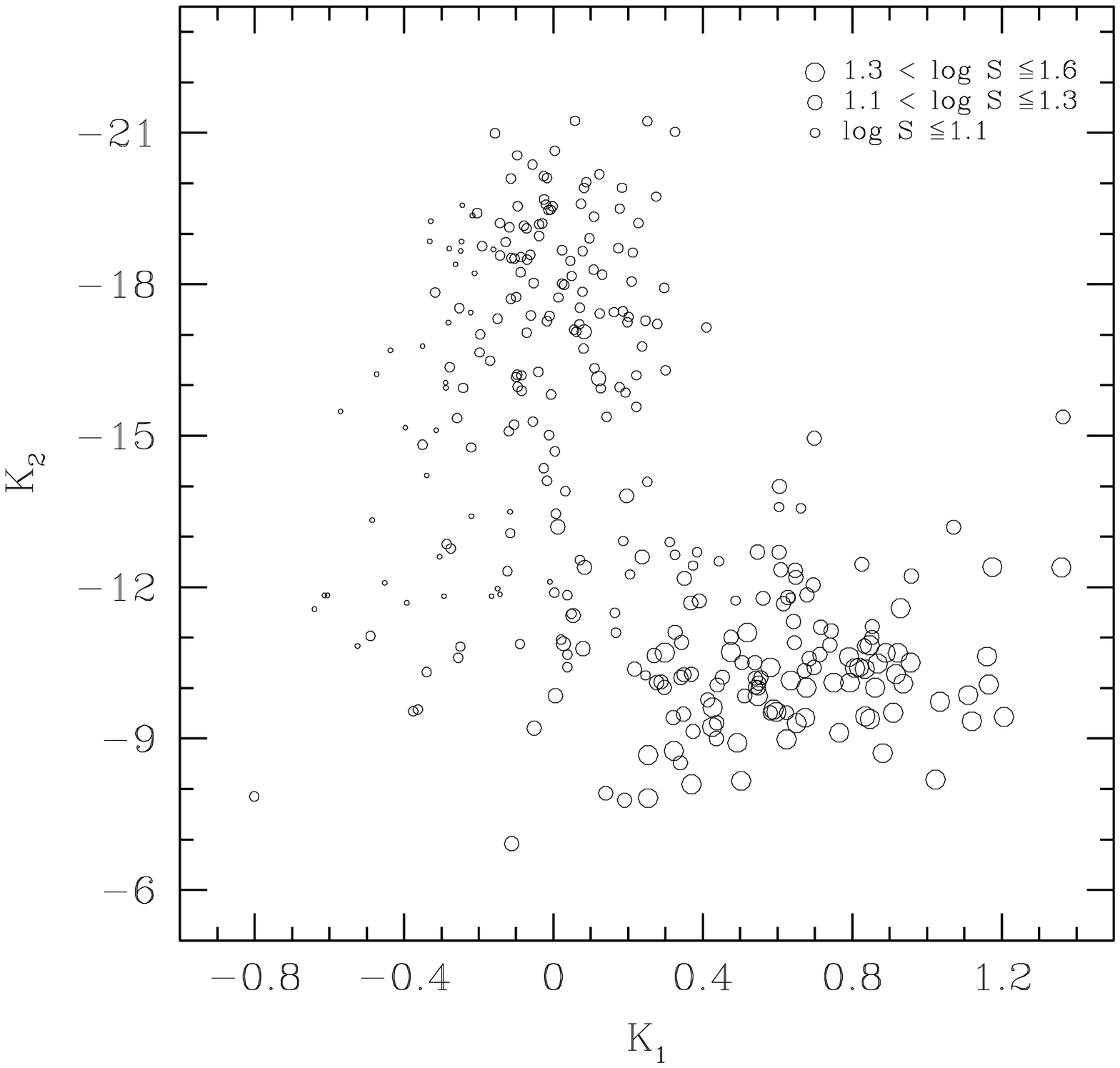}{Specific entropy of bulges and the photometric plane. The
  axes are as shown in Fig.~\ref{fa_e}.}
\section{Discussion and Conclusion}
\labsecn{discussion}  
We have compiled from the literature values of
structural parameters for bulges of galaxies of different Hubble
types, and have examined distributions of the parameters and two- and three-
dimensional correlations between  them. It may be noted
here that  the two samples of spheroidal galaxies 
in our study have been taken from dense
environments, while the systems with disks are from the field. However,
the differences between cluster and field ellipticals in their
photometric properties are marginal if not
indistinguishable, with the latter showing an increased (intrinsic)
scatter in various correlations (de Carvalho \& Djorgovski, 1992,
Gebhardt \etal 2003). Hence the inclusion of ellipticals
and dEs from clusters is justified, and the conclusions derived here
stands valid, even though the homogeneity shown by the Es
and dEs separately in our sample may partly have an environmental origin.  
Comparing the photometric structural parameters of the sample, we find that the
distributions of $n$, $\re$, $\mubzero$ and luminosity all show two
distinct peaks, with one of the peaks dominated by low luminosity
bulges, and the other by  high luminosity ones. 

The brighter bulges ($M_{\rm K} < -22$) of all Hubble types show 
similarities in various correlations like the Kormendy diagram and the
photometric plane. The fainter bulges show significant differences in 
their parameter correlations from their brighter counter parts. In 
the case of the photometric plane, while the brighter bulges show a
tight planar distribution, the fainter ones show larger scatter,
along with a deviation from the planar distribution for smaller 
$n$ values. In terms of Hubble types, ellipticals, bulges of early type
spirals and lenticulars, especially the brighter ones, form  a tight
photometric plane, while bulges of late type spirals and faint lenticulars,
and dEs, show significant deviations from the plane. 
Considering this together 
with the fact that early type (E/S0) bulges are in a 
relaxed state manifested
through their tight fundamental plane  relation (Dressler \etal 1987;
Djorgovski \& Davis 1987), we suggest here that the tight
photometric plane also reflects their relaxed nature.

We have found that the dwarf galaxies show significant similarities in
their properties with the faint end bulges of late type
spirals. Even though there are contradictory results 
  on the rotational properties of dEs, it is now becoming increasingly
  evident that in at least a significant fraction of dEs, the oblate
  structure are supported  by rotation (see Pedraz \etal 2002 and
the references therein). The bulges of late type spirals are also
known to be rotationally supported to a
greater extent than the E/S0 galaxies (see for \eg, Proctor \& Sansom
2002), suggesting that bulges of late type 
spirals could be formed by the merger of dwarfs.  This possibility needs
to be corroborated by further detailed comparison between 
bulge and disk properties of spiral galaxies and dEs.

The dichotomy in parameter distributions for bright and faint bulges
has been discussed in the literature for various Hubble
types. Kormendy (1985) noticed that bright and faint ellipsoids
differ significantly in their parameter correlations. van den Bergh
(1994) has shown that in lenticular galaxies, fainter bulges are more
prolate than the brighter ones, with a non-negligible overlap between
the two populations. Bulges of spiral galaxies also show a dichotomy,
 with later Hubble types being more elongated on an average 
than their early
type counterparts (Fathi \& Peletier 2003). These
results are consistent with the differences that we have found, in the
parameter distributions and correlations, between the bright and faint
bulges. It appears that the difference between the two kinds of 
bulges is  fundamental in nature, and is independent of the
Hubble type of the parent galaxy. But in every parameter distribution
that we have considered, there is an overlap of bright and
faint bulges, making the dividing line rather fuzzy, which helps to
account for the contradictory observations on the dichotomy (see
Graham \& Guzm\'an 2003 and the references therein).

Following the prescription of M\'arquez \etal (2000), we have obtained
the specific entropy for the bulges in our sample, using their
morphological parameters. The specific entropy for dEs and Es is more
tightly distributed than for the other Hubble types, which indicates 
greater homogeneity for these galaxies and suggests that they are at a
local maximum of the entropy, \ie of stability. The average specific
entropy shows a systematic increase as we move towards the earlier
types along the Hubble sequence, supporting  successive merging of
smaller components as in the case of hierarchical clustering (Kauffmann 
\& White 1993). This observation is in accordance with the results
obtained by  M\'arquez \etal (2000), who found that 
simulated ellipticals showed
systematic increase in their specific entropy with successive merging.
        
For the bulges of lenticulars and late type spirals, we find that the
parameter correlations are less tight  than those for Es (Figures 4 \& 5). 
This inhomogeneity is also evident in the case of the photometric plane, 
where their rms scatter is larger than that for Es. Interpreting this as 
evidence for the existence of  different scenarios of formation
for the bright and faint bulges helps to understand the observation. It is 
natural to expect bigger and brighter bulges with large $n$ and $\re$
values to be the remnants of major mergers. Khosroshahi \etal
(2004) observed a strong tendency for $n$ to systematically increase
towards the cluster center. Recent studies on luminosity functions
also suggest that there is a pronounced dominance of bright end bulges
near the central regions of clusters (de Propris \etal 2003) which is
usually attributed to the enhanced merging near the centers of
clusters occurred while the cluster is forming.

The fainter bulges, irrespective of their Hubble
class,  might have experienced fewer encounters allowing them to evolve
secularly for a longer time in the past (see Kormendy \& Kennicutt
2004 for a recent review). The existence of stable structures like
spiral arms, rings and bars are evidence for the lack of  any major
mergers/interactions in the recent past, as simulations show that
major mergers mix up the phase space rather quickly to destroy such
fine structures (see Fritze-v. Alvensleben 2004). This idea is
corroborated by the 
observation that the fine structure parameter $\Sigma$ (Schweizer
\etal 1990), which characterizes ripples, shells, plumes, boxiness,
structures \etc (in field S0s) correlates with deviations in the
Color-Magnitude diagram towards the bluer colors
(Fritze-v. Alvensleben 2004). This
implies that fainter S0s contain a statistically larger number 
of structures, signifying their lack of strong interactions.

The distribution of spiral galaxies, more abundant in the field and in
the outer regions of galaxy clusters, 
also seems to support the idea that they are likely 
to have encountered less violent
interactions. Studying the properties of disk galaxies as a function
of their isolation, Varela \etal (2004) reported that late type spirals
(Sc) are more abundant amongst isolated galaxies, while lenticulars 
are abundant in perturbed systems. Also isolated systems have a
remarkable absence of big, bright and massive galaxies, while the
opposite is true for perturbed systems.  Since the bulge properties of
early type spirals show significant similarity with those of bright Es
and S0 bulges, one might expect a significant fraction of 
early type spirals to be closer to denser environments, than their
fainter counterparts, so as to be able to experience stronger interactions.

A recent observation significant to the subject matter of this study 
is that while the fraction of Es in clusters remain almost a constant
from $z=0.5$ to $z=0$, that for spirals decreases and that for S0s
increases by the same factor of $\sim 5$(Couch \etal 1998; Fasano
\etal 2001). This has been used to suggest that a large fraction of
spirals must have transformed to lenticulars in the last $\sim5$
Gyrs. Another important factor to note about S0s is that a
significant fraction of their bulges are as bright as the
Es. As higher luminosities are attained only by major mergers, since
smaller interactions, or secular evolution for that matter, can not
increase the phase space density by more than a factor of two (see
Fritze-v. Alvensleben 2004), brighter [field] S0s might be remnants of
mergers of 
(early type) spirals, where the non vanishing or re-distributed
angular momentum preserves the presence of a stellar disk.  This
suggests a strong inter-dependence between morphological evolution and
the cosmological large scale structure formation. 
\begin{acknowledgements}
        We thank S.K. Pandey, Yogesh Wadadekar and Somak Raychaudhury for helpful comments and discussions. CDR and SB  thank  IUCAA
for the  hospitality and the use of facilities there without which
this work could not have been done. CDR thanks Centre Franco-Indien pour la Promotion de la Recherche Avancee (CEFIPRA) for a post-doctoral fellowship during which part of the work presented in the paper has been done. 

\end{acknowledgements}

\end{document}